\documentclass{aa}  

\usepackage{graphicx}
\usepackage{txfonts}
\usepackage{lipsum}
\usepackage{subcaption}       
\usepackage{hyperref}
\usepackage{gensymb}
\usepackage{xcolor}
\definecolor{Orange}{rgb}{1.0,0.5,0.15}
\definecolor{Blue}{rgb}{0,0.08,0.65}
\definecolor{Red}{rgb}{0.65,0.08,0.05}
\definecolor{Green}{rgb}{0.15,0.45,0.25}
\definecolor{Pink}{rgb}{1.0,0.05,0.5}
\definecolor{bubbles}{rgb}{0.91, 1.0, 1.0}
\definecolor{aquamarine}{rgb}{0.5, 1.0, 0.83}
\definecolor{bubblegum}{rgb}{0.99, 0.76, 0.8}
\definecolor{bluebell}{rgb}{0.74, 0.74, 0.92}
\definecolor{dollarbill}{rgb}{0.72, 0.93, 0.6}

\makeatletter
\let\linenumbers\relax

\let\modulolinenumbers\relax
\makeatother

\nolinenumbers

\begin{document}
   \title{Dissecting the Perseus-Pisces supercluster observed with CFHT-MegaCam}

    \subtitle{Exploring late-type galaxy shape 
    %correlations and \sandrine{alignments c'est suffisant non ?}
    alignments within the local cosmic web}
    
\author{
   M.~Mondelin\inst{1}
   \and
   S.~Codis\inst{1}
   \and
   J.-C.~Cuillandre\inst{1}
   \and
   R.~Paviot\inst{1}
   \and
   T.~de Boer\inst{2}
}

\institute{
   Universit\'e Paris-Saclay, Universit\'e Paris Cit\'e, CEA, CNRS, AIM, 91191 Gif-sur-Yvette, France\\
   \email{maelie.mondelin@cea.fr}
   \and
   Institute for Astronomy, University of Hawaii, 2680 Woodlawn Drive, Honolulu HI 96822, USA
}

    \date{}

\abstract{ 
Intrinsic alignments of galaxy shapes constitute a critical systematic for weak 
gravitational lensing and probe how galaxies acquire their orientations within the 
cosmic web. So far, studies have focused mostly on statistical samples of galaxies. 
In this work, we take the opposite approach and measure this signal in one local 
superstructure but with exquisite resolution up to the most external regions where 
secondary infall should be predominant. More specifically, we measure intrinsic 
alignment signals in the Perseus--Pisces Supercluster as a function of galaxy 
morphology and their radial variations into the low surface brightness regime. 

We use deep CFHT $r$-band imaging covering 367~deg$^2$ (mostly from the UNIONS survey) 
and reaching 28~mag~arcsec$^{-2}$ to measure correlation functions at three isophotal 
radii ($R_{23}$, $R_{25}$, $R_{27}$) for 2004 galaxies with $\log M_*/M_\odot > 8.6$, 
stratifying by morphological type and stellar mass in comoving coordinates.

We detect positive intrinsic alignment signals for both early- and late-type galaxies, 
extending to $\sim$1~Mpc$/h$ in comoving space, including a clear and robust signal for 
spiral galaxies. Shape-shape correlations 
($\xi_+$) are substantially stronger than position-shape correlations ($\xi_{\mathrm{g}+}$), 
with comoving measurements increasing the fraction of strongly correlated galaxies detected. 
Correlation profiles show minimal radial dependence across the three isophotes despite 
10--20\% of galaxies showing ellipticity variations. We report a robust detection 
of morphology-dependent intrinsic alignment components: late-type galaxies strongly 
dominate the shape-shape signal (86\% of strongly correlated galaxies despite representing 
69\% of the sample), residing preferentially in filamentary regions and exhibiting 
systematically higher ellipticities indicating edge-on configurations. Early-type galaxies 
show no significant over-representation in position-shape correlations, 
but concentrate near cluster/group centers where environmental enrichment becomes apparent.

The environmental segregation suggests early-types align via tidal stretching in dense 
cluster environments while late-types align via tidal torquing in filaments. These 
local-Universe measurements contribute to interpreting alignment systematics for 
upcoming surveys like \textit{Euclid}, DESI, and LSST, and highlight notably the key 
contributions of blue spiral late-type galaxies in intrinsic alignment signals at least 
in the low-redshift Universe, hence the need to model them carefully for cosmic shear 
analyses.
}


\titlerunning{The Pisces-Perseus supercluster observed with CFHT-MegaCam II}
\authorrunning{Mondelin et al.}

   \maketitle
   \section{Introduction}

In the $\Lambda$CDM cosmological framework, small matter fluctuations in the cosmic web 
drive the formation and evolution of galaxies \citep{Klypin1983, Bond1996}. This 
large-scale structure, composed of filaments, sheets, nodes, and voids, generates 
coherent tidal fields that exert torques on protogalactic gas clouds during collapse. 
This imprints preferred orientations on the angular momentum of forming galaxies 
\citep{Peebles1969, Doroshkevich1970}. Theoretical models and simulations predict that 
dark matter haloes align their spin axes and principal axes with the eigenvectors of the 
local tidal tensor, with spin directions typically perpendicular to the direction of 
strongest compression \citep{Aragon-Calvo2007, Hahn2010, Codis2012}.

The intrinsic alignment of galaxy shapes and spins constitutes both a challenge and an 
opportunity for observational cosmology. As weak gravitational lensing has emerged as a 
primary probe of dark matter and dark energy through cosmic shear measurements 
\citep{Kilbinger2015, Mandelbaum2018}, the need to disentangle intrinsic shape 
correlations from lensing-induced correlations has become critical. Intrinsic alignments 
can mimic or dilute the weak lensing signal, potentially biasing cosmological parameter 
estimates \citep{Hirata2004, Joachimi2015, Troxel2015}. In particular, tidal stretching 
of early-type galaxies by the surrounding density field has been identified as a dominant 
alignment mechanism \citep{Catelan2001, Hirata2004, Chisari2025}. We refer the readers to \cite{2024OJAp....7E..14L} for a comprehensive 
overview of alignment processes and their observational signatures. This systematic contamination has motivated extensive 
observational efforts to characterize intrinsic alignment amplitudes as a function of 
galaxy properties, redshift, and environment \citep{Mandelbaum2006, Joachimi2011, Singh2015}.

Large photometric surveys have provided the statistical power necessary to detect 
intrinsic alignments. Early measurements from the Sloan Digital Sky Survey 
(SDSS; \cite{SDSS2000, Hirata2004, Mandelbaum2006}) and the Canada–France–Hawaii 
Telescope Legacy Survey (CFHTLS; \cite{Hildebrandt2012}) revealed significant 
position-shape correlations for luminous red galaxies, confirming that massive early-type 
systems exhibit strong alignments consistent with tidal stretching by the surrounding 
density field. Stage-III surveys, including \cite{DES2023} and \cite{HSC}, have extended 
these measurements to fainter magnitudes and higher redshifts, enabling studies of 
alignment evolution and morphological dependencies. The recent UNIONS–BOSS/eBOSS 
cross-correlation has pushed these constraints further, yielding robust detections across 
3500~deg$^2$ and confirming strong luminosity scaling of alignment amplitudes 
\citep{Hervas-Peters2025}. Complementary measurements from the PAU Survey have provided 
some of the tightest constraints to date on blue galaxy alignments, finding amplitudes 
consistent with zero for star-forming systems while detecting increasing signals with 
stellar mass and luminosity for red galaxies \citep{Navarro-Girones2026}.

A parallel line of investigation has focused on the alignment of galaxy spins with the 
cosmic web filamentary structure. Hydrodynamical simulations predict a mass-dependent 
"spin flip": low-mass galaxies tend to have angular momentum vectors parallel to nearby 
filaments, consistent with tidal torquing during accretion along filaments, while massive 
galaxies exhibit perpendicular orientations, reflecting virialization and merger activity 
near filament nodes \citep{Codis2012, Dubois2014, Welker2014, Codis2015, Laigle2015}. 
Observational predictions for alignment signals as a function of environment and 
morphology further support this picture \citep{2021MNRAS.504.1694R}. This transition has 
been confirmed observationally (SAMI Galaxy Survey; \cite{Welker2020} and MaNGA 
\citep{Kraljic2021}), which provide evidence for mass-dependent and environment-dependent 
spin orientation patterns, as also suggested by theoretical arguments 
\citep{2015MNRAS.452.3369C}. Recent observations have even detected signatures of 
filament rotation itself on scales of $\sim$15~Mpc, suggesting that angular momentum 
cascades from the largest scales down to individual galaxies \citep{TudoracheJung2025}.

However, the connection between dark matter predictions and observed galaxy alignments is 
mediated by complex baryonic physics. Gas cooling, star formation, stellar and AGN 
feedback, and mergers can all suppress, enhance, or reorient alignment signals relative 
to pure dark matter expectations \citep{Chisari2015, 2018MNRAS.481.4753C}. Radiative 
cooling drives dissipative collapse that decouples stellar angular momentum from the 
underlying dark matter halo, while feedback injects energy that disrupts coherent inflows 
and randomizes orientations, particularly in low-mass systems \citep{Dubois2014, 
Zjupa2020}. Recent reviews synthesizing theoretical, simulation, and observational 
perspectives highlight that intrinsic alignments remain one of the most uncertain 
components of weak lensing systematics budgets \citep{Joachimi2015, Kiessling2015, 
Blazek2019, 2024OJAp....7E..14L}. Understanding how baryonic processes modulate 
alignment signals as a function of mass, morphology, and environment is thus essential 
both for mitigating weak lensing contamination and for constraining galaxy formation 
physics.

Looking ahead, the next generation of large-sky imaging surveys promises transformative 
advances. The Dark Energy Spectroscopic Instrument (DESI) is 
simultaneously obtaining spectra for tens of millions of galaxies, enabling environmental 
studies with unprecedented three-dimensional completeness. Meanwhile, the \textit{Euclid} space mission will provide 
complementary wide-field imaging and spectroscopy free from atmospheric distortions 
\citep{Laureijs2011, Mellier2025}. The Vera C. Rubin Observatory's Legacy Survey of Space and Time 
(LSST, \cite{Ivezic2019}) will observe billions of galaxies over 18,000~deg$^2$ with 
exquisite image quality.
\citep{DESICollaboration2016}. Early science results from \textit{Euclid} Quick 1 data 
\citep{EuclidSkyOverview, Aussel2025} already demonstrate the power of these new datasets 
for studying alignments with the cosmic web \citep{Laigle2025}. These surveys will push 
intrinsic alignment measurements to higher redshifts, lower masses, and finer 
environmental resolution than previously achievable, but their interpretation will rely 
critically on calibrations from well-characterized local samples where galaxy properties, 
morphologies, and environments can be measured with high fidelity.

In this context, the local Universe offers unique advantages for calibration studies: 
spatially resolved galaxies enabling detailed morphological classification, accurate 
three-dimensional distances for robust structure identification, and nearby superclusters 
providing high-contrast environments where tidal fields are strongest. In our companion 
study (Paper~I, \citealt{Mondelin2025}), we presented a comprehensive reconstruction of 
the Perseus–Pisces Supercluster at $z < 0.03$, identifying filaments, clusters, and 
groups using spectroscopic redshifts. This structure, containing over 3000 galaxies from 
void edges to cluster cores, provides us with a laboratory for studying how the cosmic 
web shapes galaxy properties. In this paper II, we will exploit low surface brightness features to 
probe outer stellar regions. More specifically, using LSB-optimized CFHT MegaCam imaging 
reaching 28.3~mag~arcsec$^{-2}$ (\texttt{Elixir-LSB} pipeline; \cite{Ferrarese2012}), we 
will measure galaxy shapes at multiple isophotal radii. Since outer disks are less 
gravitationally bound, they may be more susceptible to external tidal influences 
\citep{Peng2010}, allowing tests of whether alignment signatures vary with radius. For 
this reason, we will measure intrinsic alignment signals using two-point correlation 
functions, stratifying by morphology, stellar mass, and isophotal radius. The combination 
of precise morphologies, multi-radius measurements, accurate 3D structure, and 
high-contrast environment will allow us to test for alignment mechanisms in a 
well-characterized local sample.

The paper is organized as follows.
First, Section~\ref{sec:data_methods} describes the sample 
and methodology. Then, Section~\ref{sec:results} presents the intrinsic alignment measurements. Finally,
Section~\ref{sec:discussion} interprets the results in the context of theory and 
observations, and Section~\ref{sec:conclusion} concludes. Supplementary material in the appendices presents 
the validation of our correlation measurements (Appendix~\ref{app:comparisonTreecorr}), 
additional morphology--environment diagnostics (Appendix~\ref{app:morpho}), and 
morphology-dependent correlation decompositions (Appendix~\ref{app:morpho_correlations}).

\section{Data and Methods}
\label{sec:data_methods}

\subsection{Galaxy sample and shape measurements}

\begin{figure*}
    \centering
    \includegraphics[width=\textwidth]{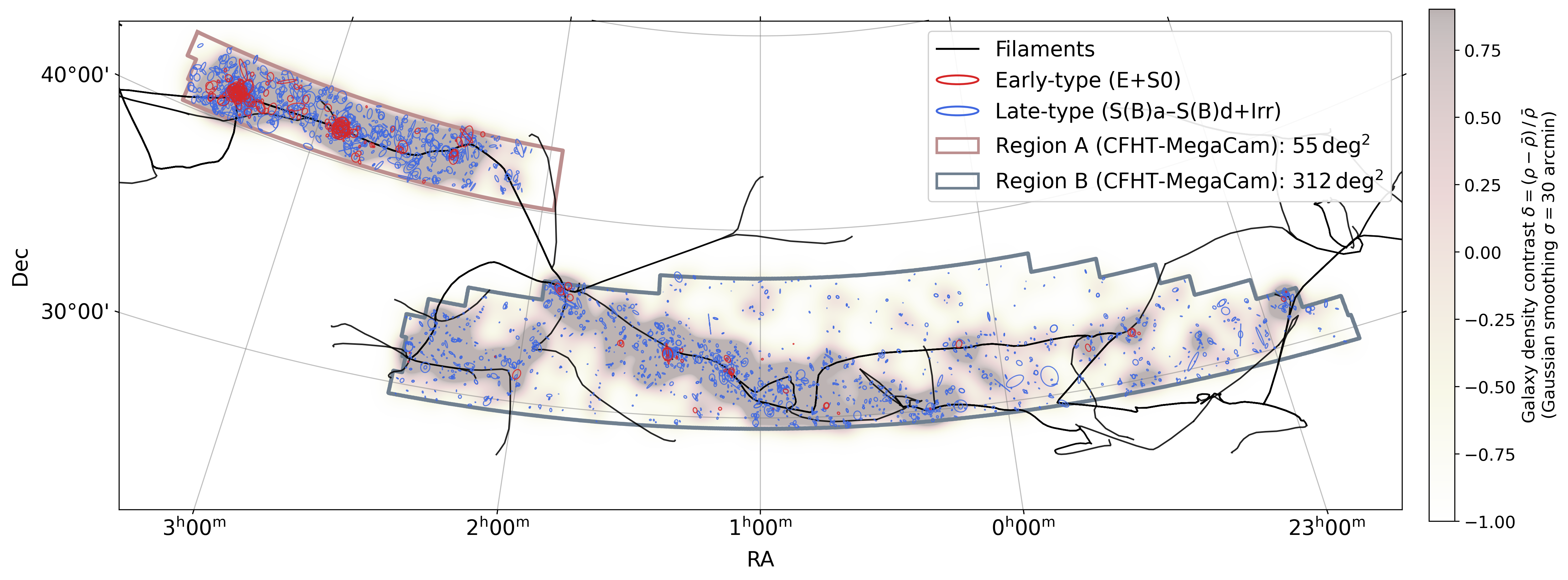}
    \caption{Spatial distribution and cosmic-web structure of the PPSC.
             Galaxy positions are shown in equatorial coordinates; background
             colour encodes the local number-density contrast
             $\delta = (\rho - \bar{\rho})/\bar{\rho}$, where $\bar{\rho}$ is
             the mean number of galaxies per bin, ranging from $\delta = -1$
             (empty regions) to several units in the densest areas.  The
             density field is computed on a 5-arcmin grid smoothed with a
             Gaussian kernel of $\sigma = 30$~arcmin.  Black curves trace
             filament spines from the projected 3D DisPerSE reconstruction
             (Paper~I).  Galaxy shapes, measured at $R_{25}$, are shown as
             oriented blue ellipses for late-type galaxies and red ellipses for
             early-type galaxies.  Regions A and B are visible as distinct
             overdensities connected by lower-density bridges. Note that region B is the UNIONS coverage of the south galactic cap \citep{Gwyn2025}.}
    \label{fig:footprint}
\end{figure*}

We analyse the galaxy sample from Paper~I: two regions, A and B, within the
Perseus–Pisces Supercluster (PPSC) observed with CFHT/MegaCam in the
$r$-band, with extensive spectroscopic redshift coverage.
Figure~\ref{fig:footprint} shows the spatial distribution in equatorial
coordinates, with background shading indicating local number density.
Filament spines extracted by DisPerSE \citep{Sousbie2011b} in Paper~I are
projected onto the RA–Dec plane and overlaid; galaxy shapes are represented
by oriented ellipses, offering a first visual impression of potential
alignments within the filamentary structure.

From the deep imaging, Paper~I extracted surface-brightness profiles using
\textsc{AutoProf} \citep{Stone2021} and \textsc{AstroPhot} \citep{Stone2023},
fitting elliptical isophotes at multiple radii.  For our intrinsic-alignment
analysis we focus on three isophotal levels: $R_{23}$, $R_{25}$, and $R_{27}$,
corresponding to surface-brightness thresholds of 23, 25, and
27~mag~arcsec$^{-2}$ in the $r$-band.  These sample progressively fainter
stellar regions from the bright inner body ($R_{23}$) through the intermediate
zone ($R_{25}$) to the extended outer envelope ($R_{27}$).

At each isophote the fit provides a semi-major axis $R_i$, semi-minor axis $b$,
and position angle $\theta_{\mathrm{PA}}$ \citep{Bartelmann2001}.  We compute
the ellipticity modulus $e = (1-q)/(1+q)$ with $q = b/R_i$, and decompose it
into Cartesian components
\begin{equation}
    e_1 = e\cos(2\theta_{\mathrm{PA}}),\qquad
    e_2 = e\sin(2\theta_{\mathrm{PA}}).
    \label{eq:e1e2}
\end{equation}
These ellipticity vectors $(e_1, e_2)$ encode both the degree of flattening and
the orientation of each galaxy, and are the fundamental observables for all
correlation measurements that follow.

\begin{figure}
    \centering
    \includegraphics[width=0.5\textwidth]{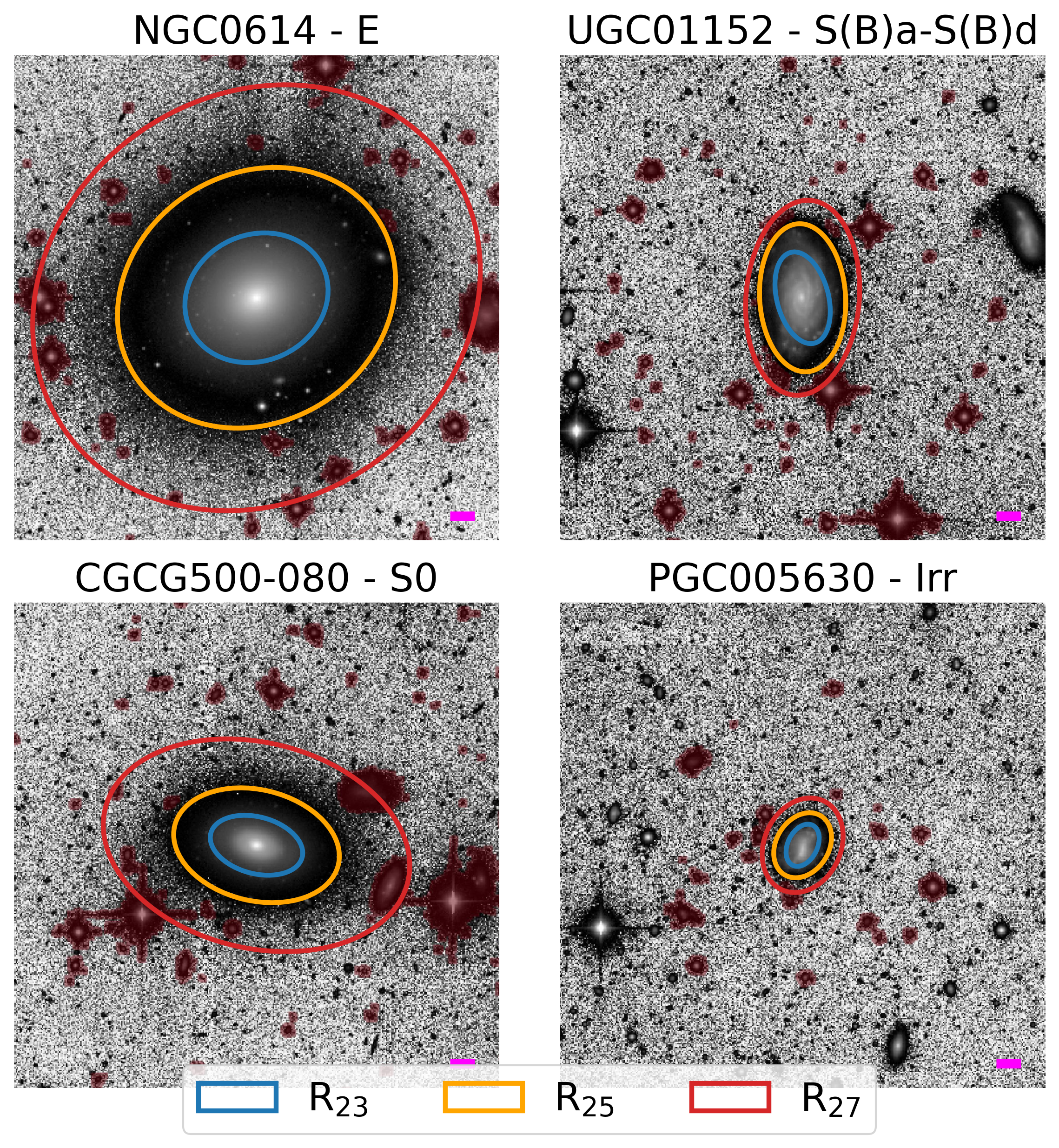}
    \caption{Isophotal shape measurements across morphological types:
             elliptical (top left), S0 (bottom left), spiral (top right),
             and irregular (bottom right).  Overlaid ellipses mark the three
             isophotal radii: $R_{23}$ (blue), $R_{25}$ (gold), and $R_{27}$
             (red), illustrating the radial evolution of galaxy shape from
             inner to outer regions.  Red-shaded areas are pixels masked during
             profile extraction.  Magenta scale bar: 10~arcsec.}
    \label{fig:morphology_radii}
\end{figure}

We adopt the morphological classifications from Paper~I, which combined visual
inspection with Sérsic-index fitting, and exclude the $\sim$10\% of galaxies
showing signs of strong tidal interactions; these systems may exhibit
anomalous alignments driven by local galaxy–galaxy torques rather than coherent
large-scale tidal fields, potentially contaminating the signal of interest.
The remaining galaxies are divided into early types (ET: ellipticals E and S0)
and late types (LT: spirals S(B)a–S(B)d and irregulars Irr).  Starting from
2843 galaxies, we apply a stellar-mass cut of $\log M_*/M_\odot > 8.5$ and
require measurements at all three isophotal radii, yielding a final sample of
2004 galaxies: 619 ET (31\%) and 1385 LT (69\%). The projected spatial distribution of the different morphological classes is
shown in Appendix~\ref{app:morpho}, which provides a visual overview of the
morphology--environment segregation across the PPSC. The mass threshold ensures
completeness above 50\% (Paper~I).  Figure~\ref{fig:morphology_radii} shows
representative examples of each morphological type with the three isophotes overlaid.

% We stratify the sample by stellar mass into two bins separated at
% $\log M_*/M_\odot = 10.0$, following \citet{Kraljic2021}.  This threshold is
% motivated by observational evidence for a transition in spin–filament alignment
% near this mass scale, with lower-mass galaxies tending to align their spin
% parallel to filaments and higher-mass galaxies showing a preferentially
% perpendicular orientation.  We note that the transition mass has been measured
% somewhat higher, $\log M_*/M_\odot \sim 10.4$–$10.9$, in the SAMI survey
% \citep{Welker2020} and in simulations \citep{2018MNRAS.481.4753C}; our choice
% is therefore a conservative lower bound on this transition, adopted for
% consistency with \citet{Kraljic2021} and to ensure sufficient statistics in
% both bins.

\subsection{Correlation functions: quantifying shape coherence}
\label{sec:corr_functions}

Intrinsic alignment manifests as statistical correlations between galaxy
orientations.  We quantify it with two complementary two-point estimators.
The \emph{shape–shape} correlation $\xi_+$ tests whether galaxies at a given
separation tend to point in similar directions.  The \emph{position–shape}
correlation $\xi_{\mathrm{g}+}$ tests whether galaxies surrounding a reference
object are preferentially oriented radially or tangentially with respect to
that reference.

Both estimators require projecting each galaxy's ellipticity onto a frame
aligned with the pair separation vector.  For galaxies $i$ and $j$ separated
by a vector with position angle $\phi$, the tangential ($e_+$) and cross
($e_\times$) components are \citep{Bartelmann2001,Troxel2015}
\begin{equation}
    e_+ = \mathrm{Re}\!\left[\varepsilon\,\mathrm{e}^{-2\mathrm{i}\phi}\right],
    \qquad
    e_\times = \mathrm{Im}\!\left[\varepsilon\,\mathrm{e}^{-2\mathrm{i}\phi}\right],
    \label{eq:e_plus}
\end{equation}
where $\varepsilon = e_1 + \mathrm{i}\,e_2$.  The shape–shape correlation is
then
\begin{equation}
    \xi_+(r) = \langle e_{+,i}\,e_{+,j}\rangle
             + \langle e_{\times,i}\,e_{\times,j}\rangle,
    \label{eq:xi_plus}
\end{equation}
averaged over all galaxy pairs at separation $r$.  Equation~\ref{eq:xi_plus}
probes intrinsic–intrinsic (II) alignment, where both galaxies have been shaped
by a common tidal field \citep{Hirata2004,Codis2015}.

For the position–shape correlation we follow the \textsc{TreeCorr} convention
\citep{Jarvis2015}: a positive value of
\begin{equation}
    \xi_{\mathrm{g}+}(r) = \langle e_{\mathrm{g}+,j}\rangle_{i},
    \label{eq:xi_gplus}
\end{equation}
where $e_{\mathrm{g}+,j} = e_{1,j}\cos 2\phi + e_{2,j}\sin 2\phi$, signals
preferential \emph{radial} elongation of surrounding galaxies toward the
reference, while a negative value indicates tangential elongation.
$\xi_{\mathrm{g}+}$ is sensitive to tidal-shear effects in which the
surrounding density field modulates the orientations of neighboring galaxies
\citep{Blazek2011}.

We compute both correlations using explicit pair counting (rather than relying
solely on \textsc{TreeCorr}), which allows us to record the individual
contributors to each separation bin.  We verified our implementation against
\textsc{TreeCorr} and found excellent agreement (Appendix~\ref{app:comparisonTreecorr}).
Separations are binned logarithmically into six bins spanning
$0.01$–$40~h^{-1}$~Mpc in comoving space.  The lower angular limit is
anchored to the typical projected galaxy size (${\sim}10~\mathrm{kpc}
\approx 0.5$~arcmin at $\langle z_{\mathrm{PPSC}}\rangle \sim 0.017$).

Throughout the paper we present results as a function of the comoving
separation $r$, computed from three-dimensional pair distances using the
spectroscopic redshifts of Paper~I. This approach naturally accounts for the
radial distribution of galaxies and ensures that pairs at different redshifts
are weighted according to their true physical separation.

Uncertainties are estimated via bootstrap resampling with $N_{\mathrm{BS}} = 500$
realisations.  In each realisation we draw, with replacement, a sample of the
same size as the original catalogue from the pool of individual galaxies,
rebuild the pair catalogue, and recompute the correlation function.  The bootstrap variance is
\begin{equation}
    \sigma_{\xi}^2 = \frac{1}{N_{\mathrm{BS}}-1}
        \sum_{k=1}^{N_{\mathrm{BS}}}
        \left(\xi^{(k)} - \bar{\xi}\right)^2,
    \label{eq:bootstrap}
\end{equation}
where $\bar{\xi}$ is the mean over all realisations.  We consider a signal
significant when it exceeds $1\sigma$.

\subsection{Multi-radius and morphology stratification}

We compute correlation functions independently at $R_{23}$, $R_{25}$, and
$R_{27}$, allowing us to test whether alignment signals vary with radius.
Inner stellar regions may be shaped primarily by baryonic processes and strong
gravitational binding, whereas outer envelopes are less bound and may more
closely trace the dark-matter halo, potentially coupling more strongly to the
large-scale tidal field \citep{Zjupa2017}.  At each radius we recalculate the
ellipticity components (Eq.~\ref{eq:e1e2}), rotate to the $(+,\times)$ frame
defined by the projected separation vector (Eq.~\ref{eq:e_plus}), and compute
$\xi_+$ and $\xi_{\mathrm{g}+}$ independently.

\begin{figure}
    \centering
    \includegraphics[width=0.48\textwidth]{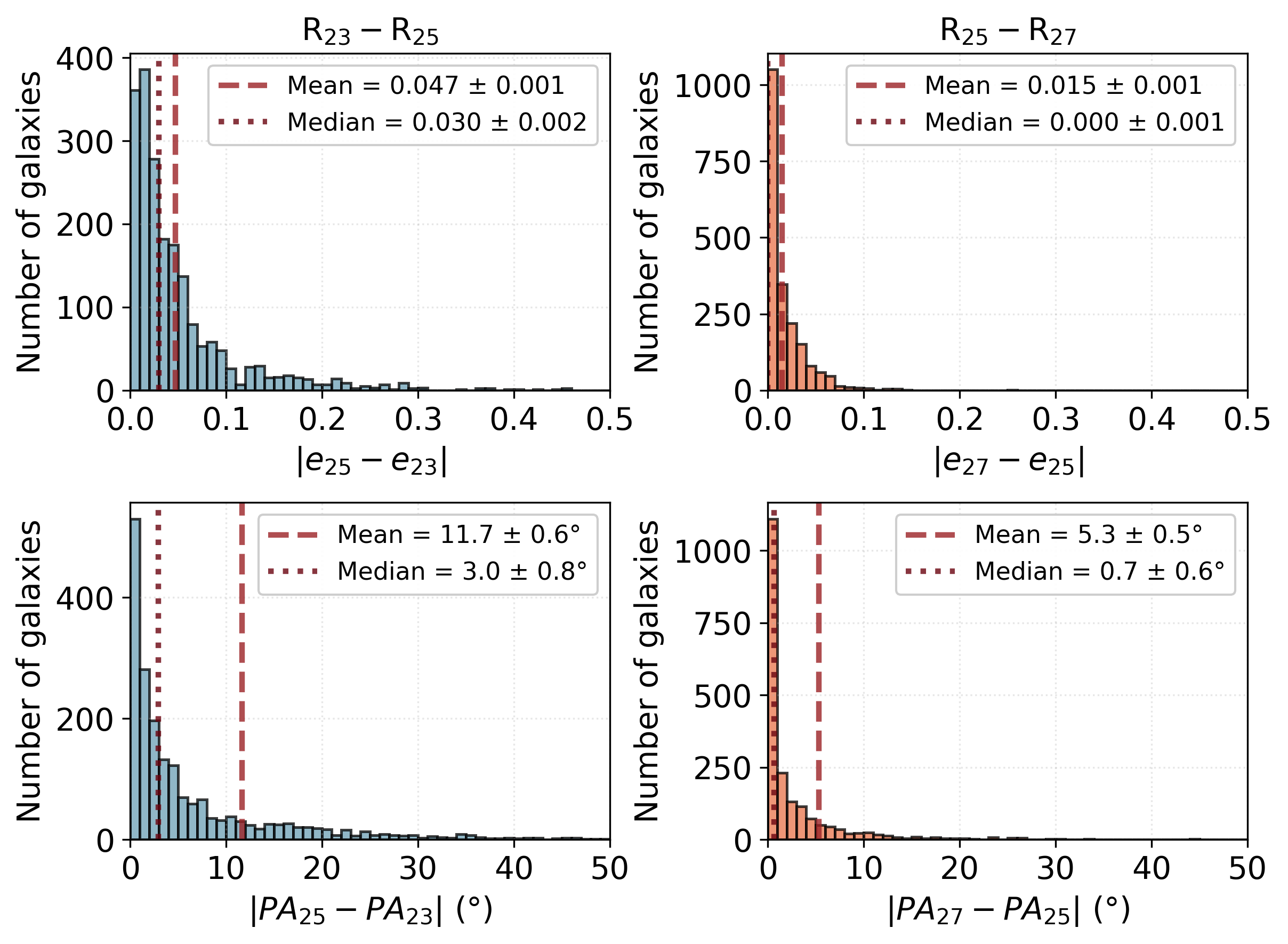}
    \caption{Variation of galaxy shapes between consecutive isophotal radii.
             Top row: ellipticity differences.  Bottom row: position-angle
             differences.  Left column: $R_{23}$ to $R_{25}$.  Right column:
             $R_{25}$ to $R_{27}$.  Dashed and dotted lines mark mean and
             median values, respectively.}
    \label{fig:deltaisophotes}
\end{figure}

Figure~\ref{fig:deltaisophotes} shows how ellipticity and position angle vary
between consecutive isophotes.  Between $R_{23}$ and $R_{25}$ the median
changes are $|\Delta e| \approx 0.030$ and $|\Delta\mathrm{PA}| \approx 3.0\degr$,
with 31\% of galaxies exceeding $|\Delta e| > 0.05$ and 37\% exceeding
$|\Delta\mathrm{PA}| > 5\degr$.  From $R_{25}$ to $R_{27}$ these variations
decrease to median values of $0.010$ and $1.5\degr$, with only 8\% and 17\%
exceeding the respective thresholds. $\sim$90\% galaxies therefore maintain coherent
shapes across radii.  Both distributions nonetheless show extended tails,
particularly in position angle (mean $12.0\degr$ from $R_{23}$ to $R_{25}$,
well above the median), driven by outliers with isophotal twists exceeding
$20\degr$; visual inspection confirms these are predominantly late-type spirals
whose central bars or asymmetric arms create radial shape variations.

\subsection{Direct alignment with the cosmic web}
\label{sec:method_direct_align}

\begin{figure}
    \centering
    \includegraphics[width=0.48\textwidth]{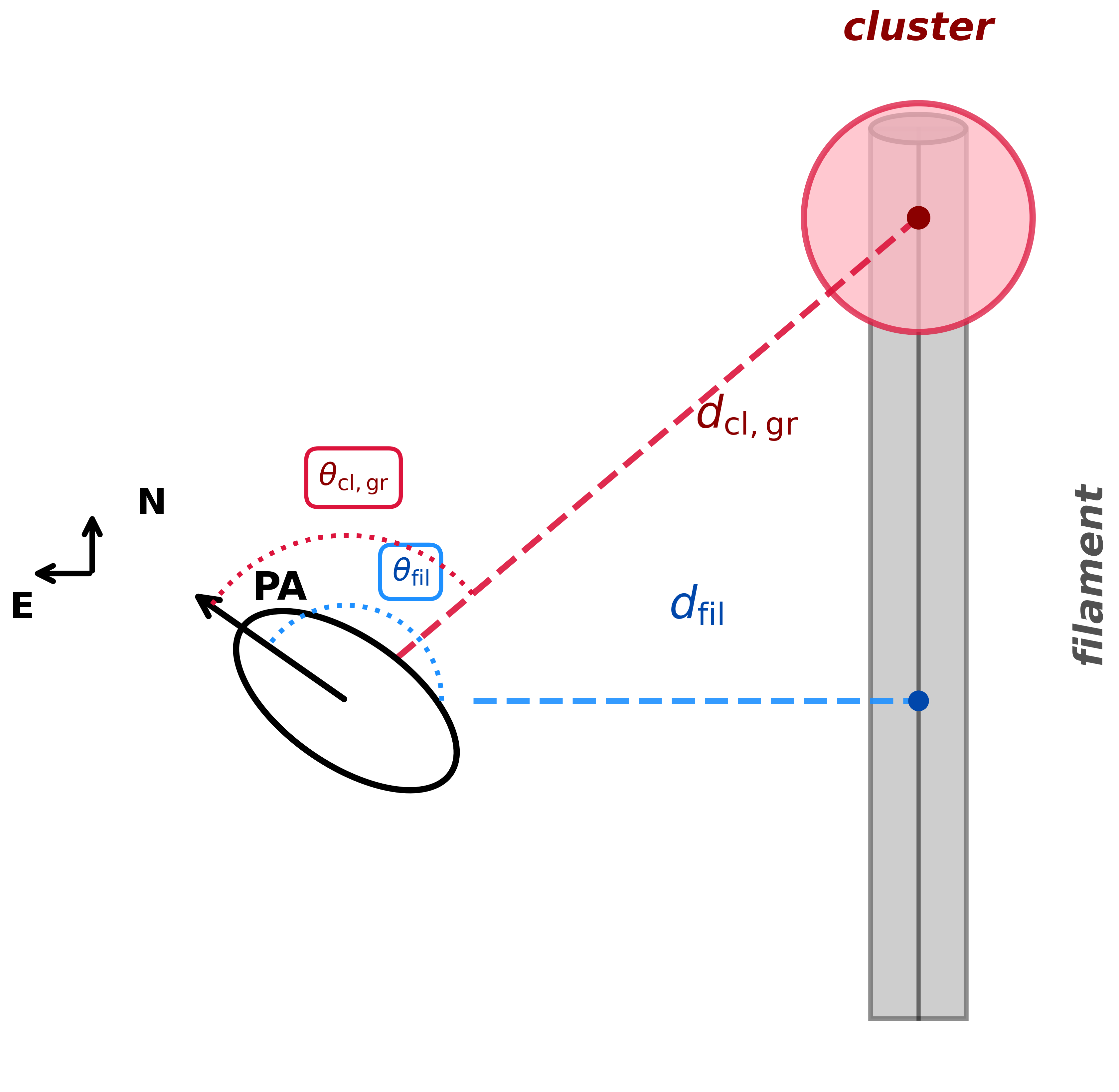}
    \caption{Schematic illustration of alignment measurements and environmental 
             parameters. The filament 
             is shown as a grey cylinder. A galaxy (black ellipse) is characterized by 
             its position angle PA, its distance to the nearest filament spine 
             ($d_{\rm fil}$, blue dashed line), and its distance to the nearest 
             cluster or group center ($d_{\rm cl,gr}$, red dashed line). The two angles $\theta_{\rm fil}$ and $\theta_{\rm cl,gr}$ are also represented.}
    \label{fig:alignment_scheme}
\end{figure}

For each galaxy we define the alignment angle $\theta_{\mathrm{cl/fil}} \in [0\degr, 90\degr]$ as the angle, projected on the plane of the sky, between the galaxy
major axis and a reference cosmic-web direction. Figure~\ref{fig:alignment_scheme} 
illustrates the geometric setup and the key environmental parameters used throughout 
this analysis.

The major-axis unit vector in the (East, North) frame is
\begin{equation}
    \hat{a} = (\sin\mathrm{PA},\;\cos\mathrm{PA}),
    \label{eq:major_axis}
\end{equation}
where PA is the position angle measured from North toward East.  

We compute two alignment angles for each galaxy, corresponding to two different
cosmic-web directions. For $\theta_{\mathrm{fil}}$, the reference direction is
the local tangent to the nearest DisPerSE filament spine: we project
each skeleton segment onto the sky, identify the node closest to the galaxy,
and compute the tangent using the two adjacent nodes, applying the standard
$\cos\mathrm{Dec}$ correction for the convergence of meridians. For
$\theta_{\mathrm{cl}}$, the reference direction is the projected line of sight
from the galaxy to the centre of its host cluster or group, defined by
the position of the central galaxy (Paper~I). Because $\hat{a}$ and $-\hat{a}$ are equivalent, we use the absolute dot
product:
\begin{equation}a
    \theta_{\mathrm{cl/fil}} = \arccos\!\left(|\hat{a}\cdot\hat{r}|\right),
    \label{eq:theta_alignment}
\end{equation}
where $\hat{r}$ is either the filament tangent or the cluster-centric direction.
A value $\theta_{\mathrm{cl/fil}} < 45\degr$ indicates that the major axis tends to lie along the structure; $\theta_{\mathrm{cl/fil}} > 45\degr$ indicates alignment
perpendicular to it. The isotropic reference is
$\langle\theta_{\mathrm{cl/fil}}\rangle = 45\degr$ (dashed line in
Fig.~\ref{fig:app_alignment_summary}).

Measurements are performed at each of the three isophotal radii ($R_{23}$, $R_{25}$, $R_{27}$). We present results for $R_{25}$ throughout the main text; results for $R_{23}$ and $R_{27}$ are consistent.
Following the environmental classification of Paper~I, each galaxy is assigned
to one of four environments based on its distance to the nearest cluster or
group centre ($d_{\mathrm{gr,cl}}$) and to the nearest filament spine
($d_{\mathrm{fil}}$). By comparing the environmental distributions of strongly
correlated galaxies with those of the full sample, we test whether enhanced
alignment is associated with specific large-scale structures.

\section{Results}
\label{sec:results}

We present measurements of intrinsic alignment signals in the PPSC. We first
analyse correlation functions computed for the full galaxy sample, then
investigate which galaxies contribute most strongly to the detected signals,
and finally connect alignment patterns to the large-scale structure
environment characterised in Paper~I.

% -----------------------------------------------------------------------
\subsection{Detection of intrinsic alignment signals}
\label{subsec:correlation_results}
% -----------------------------------------------------------------------

\begin{figure}
    \centering
    \includegraphics[width=0.495\textwidth]{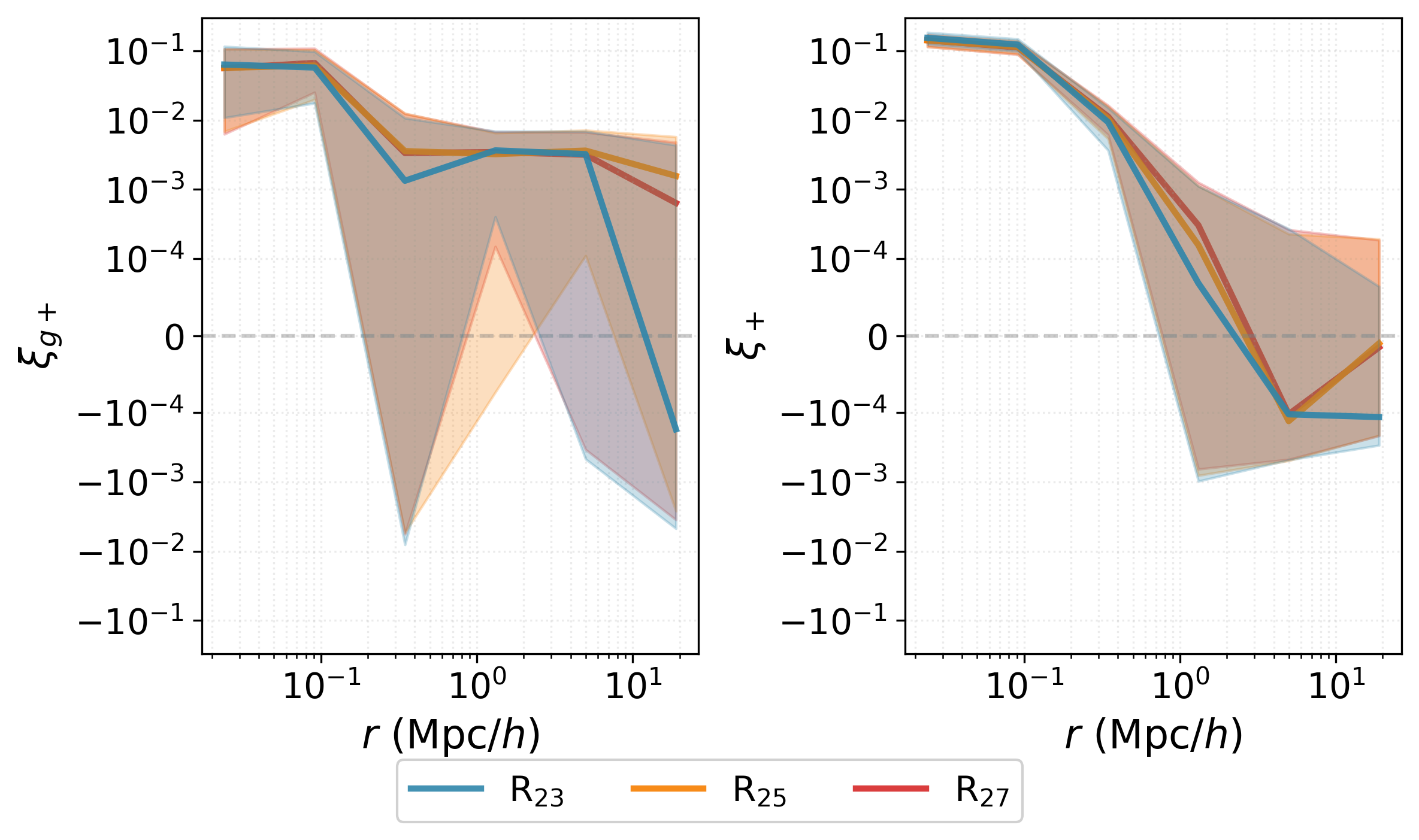}
    \caption{Two-point correlation functions for the full galaxy sample as a
             function of comoving separation $r$. Left panel: position–shape
             correlation $\xi_{\mathrm{g}+}$. Right panel: shape–shape
             correlation $\xi_+$. Three isophotal radii are shown: $R_{23}$
             (blue), $R_{25}$ (orange), $R_{27}$ (red). Shaded regions indicate
             bootstrap uncertainties.}
    \label{fig:corrtot}
\end{figure}

Figure~\ref{fig:corrtot} shows the correlation functions measured for the full
galaxy sample as a function of comoving separation $r$. Both $\xi_{\mathrm{g}+}$
and $\xi_+$ exhibit a clear positive signal at small separations. The
position–shape correlation $\xi_{\mathrm{g}+}$ shows a non-zero large-scale
amplitude of order ${\sim}4\times10^{-3}$, with a significant enhancement at
$r \sim 0.09~h^{-1}$~Mpc (S/N~$\sim3$). The shape–shape correlation $\xi_+$
remains consistent with zero on large scales while showing a strong detection
at $r \lesssim 0.1~h^{-1}$~Mpc (S/N~$\sim10$).

The signal at $r \lesssim 0.1~h^{-1}$~Mpc probes the one-halo regime, where
galaxy pairs reside within the same dark matter halo. At these scales,
alignments are driven by the local tidal field internal to the halo, naturally
producing the strong signal observed. This scale matches the typical size of
galaxy groups and clusters in the two PPSC regions (Paper~I), supporting the
interpretation that the dominant contribution arises from galaxies sharing the
same gravitational potential.

At larger separations ($r \gtrsim 1~h^{-1}$~Mpc), $\xi_{\mathrm{g}+}$ settles
toward a roughly constant amplitude (${\sim}4\times10^{-3}$), marking the
transition to the two-halo regime in which galaxy pairs belong to distinct halos
correlated through the large-scale tidal field, for example galaxies in
neighbouring groups connected by filaments. This amplitude is consistent with
results from large spectroscopic surveys reporting values of
${\sim}10^{-3}$--$10^{-2}$ at $r \sim 5$--$100~h^{-1}$~Mpc for the SDSS LOWZ
sample \citep{singh2024}.  The three isophotal radii yield broadly similar
profiles although we observe a systematic trend for the large-scale amplitude to (slightly) increase
with increasing $R_{\mathrm{iso}}$, indicating that outer isophotes are indeed
more sensitive to large-scale tidal distortions.
% \sandrine{peut etre souligner que dans tous les plots, ce plateau a l'air plus
% haut à grand $R_{iso}$ ce qui va dans le sens attendu} --> done

Decomposing these correlation functions by morphological type 
(see also Appendix~\ref{app:morpho_correlations} for more details) reveals that the detected signals 
seem to be dominated by the late-type population in first approximation. 
This morphological dichotomy motivates the detailed analysis of strongly 
correlated galaxies presented in Sect.~\ref{subsec:dominant_galaxies}.

% -----------------------------------------------------------------------
\subsection{Strongly aligned galaxies}
\label{subsec:dominant_galaxies}
% -----------------------------------------------------------------------

\subsubsection{Identifying strongly correlated galaxies}

Detecting an average positive correlation does not reveal whether all galaxies
are weakly aligned or whether a minority drives the signal. To distinguish
these scenarios, we identify galaxies contributing disproportionately to the
measured correlations.

For each separation bin where $\xi$ is significantly positive (first three bins
in comoving space), we compute for each galaxy $i$ its cumulative contribution
\begin{equation}
    C_i = \sum_{j \in \mathrm{bin}}
          \left(e_{t,i}\,e_{t,j} + e_{\times,i}\,e_{\times,j}\right).
    \label{eq:contribution}
\end{equation}
Galaxies with $C_i > \langle C\rangle + \sigma_C$ are flagged as
``strongly correlated'' (SCG hereafter). These are systems whose orientations are
unusually coherent with their neighbours. The selection is performed
independently for each correlation function ($\xi_+$ and
$\xi_{\mathrm{g}+}$, in comoving coordinates) and isophotal radius ($R_{23}$,
$R_{25}$, $R_{27}$).

\subsubsection{Global properties of strongly correlated galaxies}
\label{subsec:dominant_galaxies_results}

Table~\ref{tab:fractions_R25} quantifies the SCG populations at $R_{25}$.
For $\xi_{\mathrm{g}+}(r)$, SCGs represent 19.4\% of the whole population,
split into 6.0\% early-type and 13.4\% late-type.

\begin{table}[htbp]
\centering
\caption{Fractions of strongly correlated galaxies (SCG) at $R_{25}$.
         The full sample contains 2017 galaxies: 30.7\% early-type (ET)
         and 69.3\% late-type (LT).}
\label{tab:fractions_R25}
\begin{tabular}{lccc}
\hline\hline
Function & Total (\%) & ET (\%) & LT (\%) \\
\hline
$\xi_{\mathrm{g}+}(r)$ & $19.4 \pm 0.9$ & $6.0 \pm 0.5$ & $13.4 \pm 0.8$ \\
$\xi_+(r)$              & $12.0 \pm 0.7$ & $1.7 \pm 0.3$ & $10.3 \pm 0.7$ \\
\hline\hline
\end{tabular}
\end{table}

For shape–shape alignments $\xi_+(r)$, late-types clearly dominate, representing
${\sim}86\%$ of SCGs ($10.3\%/12.0\%$), a strong enrichment relative to their
baseline fraction of 69.3\%. For position–shape alignments $\xi_{\mathrm{g}+}(r)$,
the morphological fractions among SCGs (31\% ET, 69\% LT) mirror the baseline
composition of the sample, indicating that morphology alone does not predict
selection into the $\xi_{\mathrm{g}+}$ SCG population.

This contrast between the two estimators likely reflects a competition between
ellipticity and environment. For $\xi_{\mathrm{g}+}$, the signal scales as
$e \times \delta$: the density bias toward dense environments favours early-types,
while the ellipticity factor favours late-types, and the two effects approximately
cancel, leaving the morphological mix unchanged. For $\xi_+$, the signal scales
as $e^2$, strongly weighting the more elongated late-type population regardless
of environment.

Regarding the overlap between selections, about 39\% of SCGs appear in both
$\xi_{\mathrm{g}+}(r)$ and $\xi_+(r)$, while 46\% appear only in
$\xi_{\mathrm{g}+}(r)$ and 15\% only in $\xi_+(r)$. This partial overlap
is consistent with the interpretation that position–shape correlations are
more sensitive to galaxies residing in dense environments where tidal fields
are strongest, while shape–shape correlations preferentially select highly
elongated galaxies whose shapes are coherent over larger volumes.

In this picture, $\xi_{\mathrm{g}+}$ is likely dominated by central–satellite
pairs: the position of the central galaxy traces the local overdensity, while
its shape is radially stretched by the halo potential, producing a strong
density–shape signal on the one-halo scale. This is consistent with findings in
the Horizon-AGN simulation, where spheroidal galaxies radially align toward
overdensities and toward each other, and where satellites contribute
significantly to the alignment signal through their preferential orientation
around central galaxies \citep{Chisari2015,Chisari2017}. By contrast,
$\xi_+$ mixes pairs at all halo-centric positions and is dominated by the
more numerous satellite–satellite population; its signal is therefore more
sensitive to the global elongation of the galaxy population than to the local
density environment. The preferential distribution of satellites within the
galactic plane of massive red centrals \citep{Welker2018} further reinforces
the connection between the one-halo alignment signal and the large-scale tidal
field that shapes both the central and its satellite system.

Table~\ref{tab:masses_R25} presents mean stellar masses of SCGs at $R_{25}$.
Strongly correlated galaxies show at most modest mass enhancements of
$0.03$--$0.06$~dex relative to the full sample, consistent within the
measurement uncertainties for both morphological types and both correlation
functions. Stellar mass therefore does not appear to be a primary driver of
alignment strength in our sample; the morphological dichotomy described above
is more naturally attributed to structural (ellipticity) and environmental
effects.

\begin{table}[htbp]
\centering
\caption{Mean stellar masses of SCGs at $R_{25}$.}
\label{tab:masses_R25}
\begin{tabular}{lcc}
\hline\hline
Sample & \multicolumn{2}{c}{$\langle\log M_*/M_\odot\rangle$} \\
 & Early-type & Late-type \\
\hline
All galaxies            & $10.39 \pm 0.02$ & $9.74 \pm 0.02$ \\
$\xi_{\mathrm{g}+}(r)$ & $10.43 \pm 0.04$ & $9.79 \pm 0.03$ \\
$\xi_+(r)$              & $10.43 \pm 0.06$ & $9.78 \pm 0.04$ \\
\hline\hline
\end{tabular}
\end{table}

\subsubsection{Morphological properties of strongly correlated galaxies}

\begin{figure*}
    \centering
    \includegraphics[width=\textwidth]{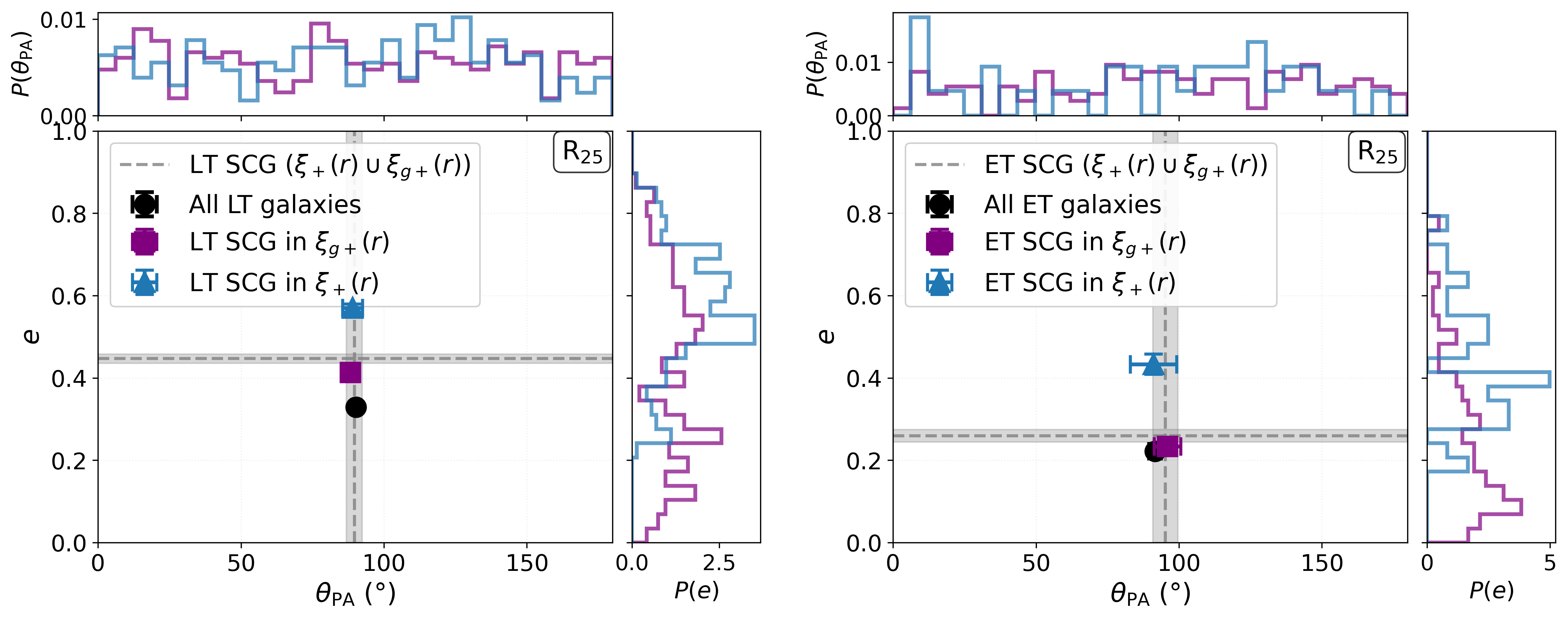}
    \caption{Morphological properties of SCGs in the ellipticity–position
             angle plane at $R_{25}$: late-type (left) and early-type (right).
             Black circles show the mean ellipticity and position angle of all
             galaxies within each morphological class. Colored symbols show the
             mean positions of SCGs selected by $\xi_{\mathrm{g}+}(r)$ (purple
             squares) and $\xi_+(r)$ (blue triangles). Gray dashed lines and
             shaded regions show the mean and uncertainty of the union of all
             SCGs ($\xi_+(r) \cup \xi_{g+}(r)$). Top and right panels show
             marginal distributions $P(\theta_{\rm PA})$ and $P(e)$ for each
             SCG population.}
    \label{fig:ellip_pa}
\end{figure*}

Figure~\ref{fig:ellip_pa} shows the distribution of SCGs in the
ellipticity–position angle plane at $R_{25}$.

\begin{table}
\centering
\caption{Ellipticity enhancement ratios $\langle e|SCG \rangle / \langle e \rangle$ 
         for SCGs selected by $\xi_{\mathrm{g}+}(r)$ and $\xi_+(r)$ at $R_{25}$. 
         Expected ratios for linear ($1+\sigma/\mu$) and quadratic ($1+2\sigma/\mu$) 
         scaling are shown for comparison.}
\label{tab:ratio_e}
\begin{tabular}{lcccc}
\hline\hline
Morphology & $\langle e \rangle$ & $\xi_{\mathrm{g}+}$ & $\xi_+$ & Linear / Quadratic \\
\hline
Late-type  & 0.33 & 1.25 & 1.73 & 1.24 / 1.48 \\
Early-type & 0.22 & 1.05 & 1.95 & 1.34 / 2.68 \\
\hline\hline
\end{tabular}
\end{table}

For late-types, the most prominent feature is a systematic offset toward higher
ellipticities. The full sample has mean $\langle e \rangle \approx 0.33$, while 
SCGs reach $\langle e \rangle_{\mathrm{SCG}} \approx 0.41$--$0.57$ depending on 
the selection, with $\xi_+$-selected galaxies showing the highest values. 
Table~\ref{tab:ratio_e} quantifies this enhancement: the ratio 
$\langle e \rangle_{\mathrm{SCG}} / \langle e \rangle_{\mathrm{all}}$ is 
$1.25$ for $\xi_{\mathrm{g}+}$ and $1.73$ for $\xi_+$ at $R_{25}$. 

The ratio $\langle e|SCG \rangle / \langle e \rangle$ allows us to distinguish 
between two scenarios: purely statistical enrichment due to the $e$ or $e^2$ 
weighting of the estimators, versus genuine physical alignment. If the 
enrichment were purely statistical, we would expect ratios of $1+\sigma/\mu \approx 1.24$ 
for linear ($e$) weighting and $1+2\sigma/\mu \approx 1.48$ for quadratic ($e^2$) 
weighting in late-types, where $\sigma/\mu \approx 0.24$ is the coefficient of 
variation of the ellipticity distribution. The observed values ($1.25$ for 
$\xi_{\mathrm{g}+}$ and $1.73$ for $\xi_+$) bracket and slightly exceed these 
predictions, confirming the expected $e$ and $e^2$ sensitivity. The modest 
excess beyond pure statistical expectations for $\xi_{\mathrm{g}+}$ suggests 
a mixed signal: genuine alignment of disk orientations combined with a contribution 
from more edge-on systems (which appear more elliptical in projection). For $\xi_+$, 
the strong enrichment reflects the dominance of edge-on disks, which carry both 
higher ellipticities and stronger shape correlations. Position angles remain 
clustered near $90°$--$95°$ across all selections, with no significant systematic 
offset relative to the full sample, suggesting that the correlation functions are 
primarily sensitive to the degree of elongation rather than to a preferred 
orientation on the sky.

For early-types ($\langle e \rangle \approx 0.22$ for the full subsample), 
the contrast between the two estimators is striking. $\xi_{\mathrm{g}+}$-selected 
SCGs have ellipticities consistent with the full population 
($\langle e \rangle_{\mathrm{SCG}} \approx 0.23$, ratio $1.05$), indicating that 
the density–shape signal is driven by coherent orientations rather than by a 
selection of intrinsically more elliptical galaxies. This is a "clean" alignment 
signal, independent of ellipticity. In contrast, $\xi_+$-selected SCGs show 
markedly higher values ($\langle e \rangle_{\mathrm{SCG}} \approx 0.43$, ratio $1.95$), 
approaching the $1+2\sigma/\mu \approx 2.68$ expectation for quadratic scaling. 
This reveals that the shape–shape correlation for early-types is dominated by 
galaxies that have been most strongly deformed by the tidal field. Ellipticity 
becomes essential for early-types to contribute to $\xi_+$, while their contribution 
to $\xi_{\mathrm{g}+}$ is driven primarily by their biased clustering in dense 
environments. These conclusions are robust to the choice of isophotal radius.

In summary: for late-types, both estimators show ellipticity enrichment, reflecting 
the importance of viewing angle (edge-on disks are both more elliptical and better 
aligned with large-scale structure). For early-types, i) $\xi_{\mathrm{g}+}$ is mostly independent of ellipticity and trace mostly the special location of elliptical galaxies in dense environment and their radial orientation, while ii)
$\xi_+$ shows some ellipticity-dependence hence mixing tidal deformation (which increases 
ellipticity) with orientation coherence.

\subsubsection{Environmental distribution of strongly correlated galaxies}
\label{subsec:environment_analysis}

\begin{figure*}
    \centering
    \includegraphics[width=\textwidth]{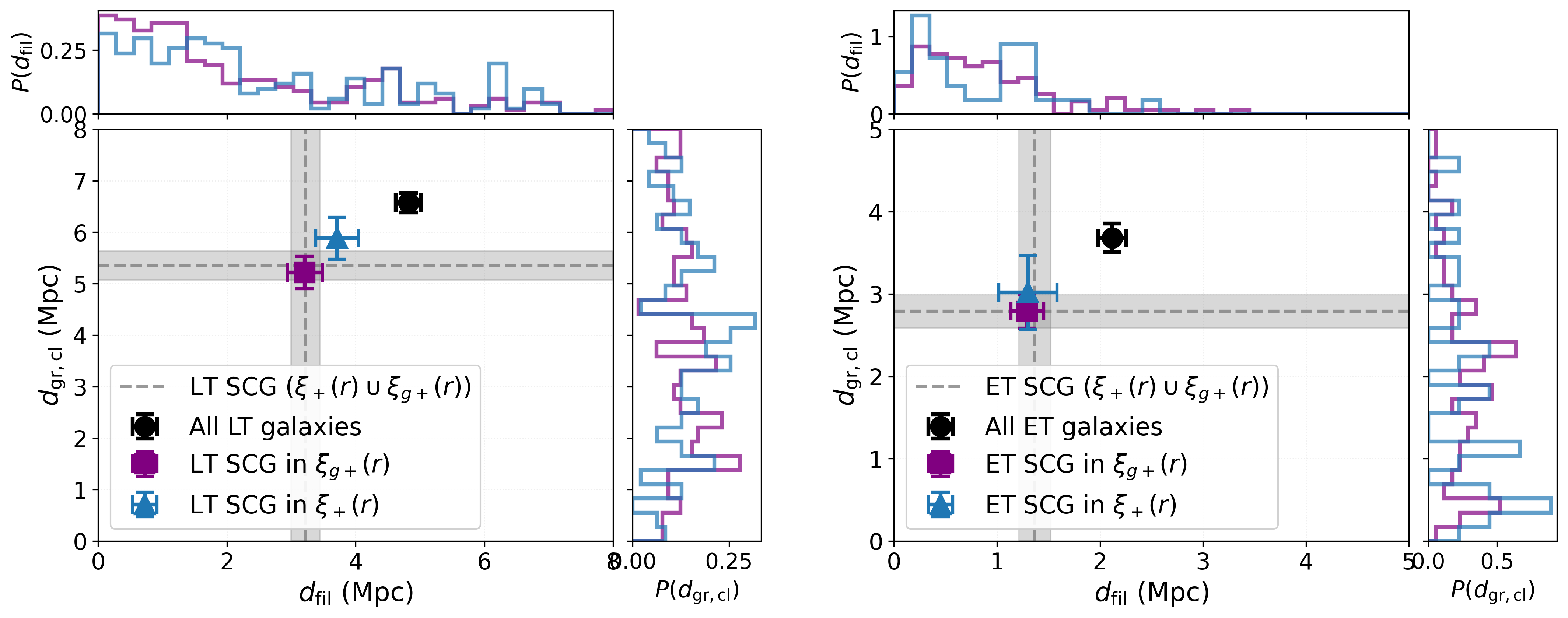}
    \caption{Mean environmental positions of SCGs with shapes measured at
             $R_{25}$: distance to the nearest filament ($x$-axis) vs.\
             distance to the nearest cluster or group ($y$-axis) for late-type
             (left) and early-type (right) galaxies. Black circles show the
             mean positions of all galaxies within each morphological class.
             Colored symbols show the mean positions of SCGs selected by
             $\xi_{\mathrm{g}+}(r)$ (purple squares) and $\xi_+(r)$ (blue
             triangles). Gray dashed lines and shaded regions show the mean
             and uncertainty of the union of all SCGs ($\xi_+(r) \cup \xi_{g+}(r)$).
             Top and right panels show marginal distributions $P(d_{\rm fil})$
             and $P(d_{\rm gr,cl})$ for each SCG population.}
    \label{fig:distance}
\end{figure*}

Figure~\ref{fig:distance} shows the environmental positions of SCGs,
characterised by their distances to the nearest filament ($d_{\mathrm{fil}}$)
and nearest cluster or group ($d_{\mathrm{gr,cl}}$). Results are consistent
across the three isophotal radii.

For late-types, the full sample has mean filament distance ${\sim}5.0$~Mpc
and cluster distance ${\sim}6.5$~Mpc. Strongly correlated late-types are
systematically closer to filaments (mean ${\sim}3.3$~Mpc, an offset
of ${\sim}2$~Mpc) and somewhat closer to clusters
(${\sim}5.5$~Mpc, offset ${\sim}1$~Mpc).

For early-types, the full subsample already lies closer to both structures
(${\sim}2.2$~Mpc from filaments, ${\sim}3.7$~Mpc from clusters)\footnote{This
is expected since late-type galaxies are more numerous in smaller filaments not extracted in the coarse DisPerSE reconstruction used here.}. Strongly
correlated early-types are found ${\sim}0.5$~Mpc closer to both filaments and
clusters than the full early-type sample.
 
\begin{figure}
\centering
\includegraphics[width=\linewidth]{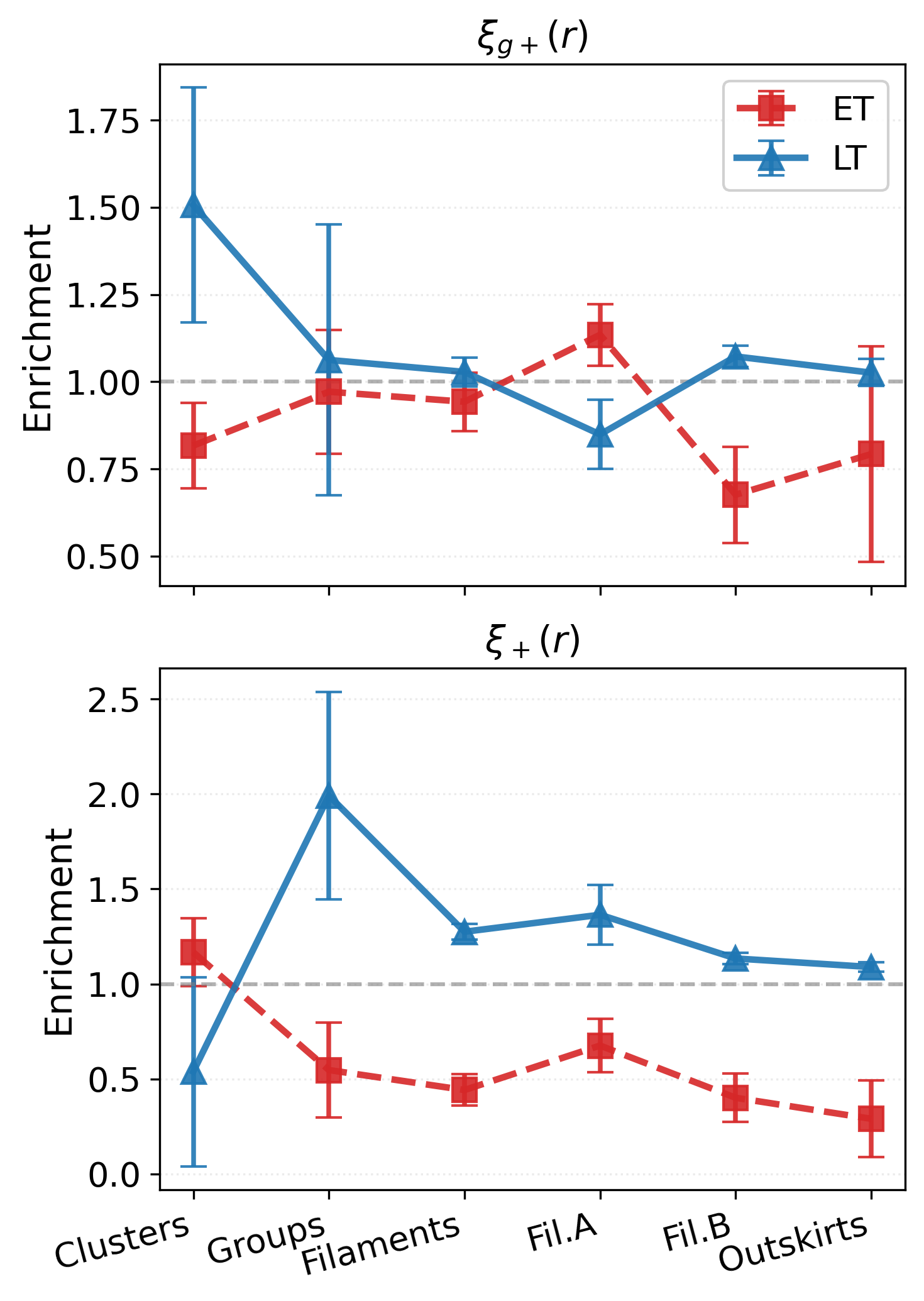}
\caption{Morphological enrichment of strongly correlated galaxies (SCGs) as a
         function of environment at $R_{25}$, for $\xi_{\mathrm{g}+}(r)$ (top)
         and $\xi_+(r)$ (bottom). The enrichment is defined as the fraction of
         a given morphological type among SCGs divided by its fraction in the
         full sample within the same environment. A value above unity indicates
         over-representation; below unity indicates under-representation.
         Early-type (ET) galaxies are shown in red; late-type (LT) galaxies
         in blue.}
\label{fig:env_enrichment}
\end{figure}

Figure~\ref{fig:distance} shows that SCGs are systematically found closer to
filaments and clusters than the full sample, but does not reveal whether this
environmental preference depends on morphology. Figure~\ref{fig:env_enrichment}
addresses this directly by showing the enrichment of early-type and late-type
galaxies among SCGs relative to their baseline fractions, as a function of
large-scale environment.
 
For $\xi_{\mathrm{g}+}(r)$, the morphological mix of SCGs is close to that of
the full sample across all environments, confirming the result of
Sect.~\ref{subsec:dominant_galaxies_results} that morphology alone does not
drive selection into the position--shape SCG population. There is however a
mild tendency for early-type SCGs to be overrepresented in clusters and groups, where the early-type enrichment rises modestly above unity, while late-type
enrichment remains near unity and flat across environments. This is consistent with the picture in which $\xi_{\mathrm{g}+}$ is sensitive to the local density field, which early-types preferentially trace in the densest structures.
 
For $\xi_+(r)$, the picture is strikingly different. Late-type galaxies are
strongly overrepresented among SCGs in groups, filaments, and outskirts, with
enrichment well above unity, while early-type galaxies fall significantly below
unity in all environments except clusters, where both morphologies show comparable
enrichment near unity. This depletion of early-type SCGs in underdense environments
is especially pronounced: outside clusters, early-type SCGs are rare while
late-type SCGs remain abundant. This environmental pattern confirms that
shape--shape alignments are driven by a population of coherently elongated
late-type galaxies distributed throughout the large-scale structure, rather
than by galaxies concentrated in the densest nodes.

% -----------------------------------------------------------------------
\subsection{Direct alignments with the cosmic web}
\label{subsec:direct_alignments}
% -----------------------------------------------------------------------

To investigate further the link between intrinsic alignments and the environment, we now study the direct alignments of our sample of galaxies with the cosmic web.

\begin{figure*}
    \centering
    \includegraphics[width=0.95\textwidth]{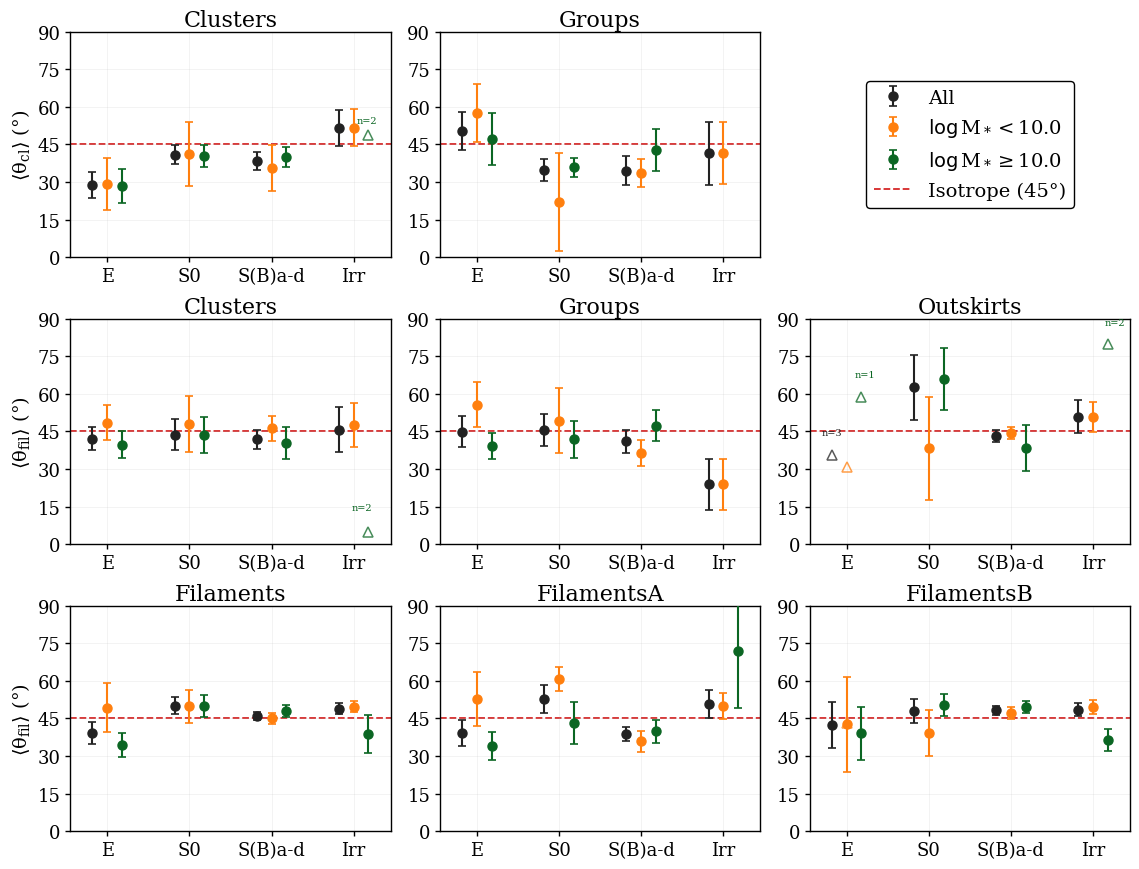}
    \caption{Median alignment angle between the projected semi-major axis and
the cosmic web at $R_{25}$, as a function of morphological type
(E, S0, S(B)a--d, Irr) and large-scale environment. Top row:
angle with the direction to the nearest cluster or group centre
($\langle\theta_{\mathrm{cl}}\rangle$) for clusters and groups.
Middle row: angle with the nearest filament tangent
($\langle\theta_{\mathrm{fil}}\rangle$) for clusters, groups, and
outskirts. Bottom row: same for filaments, filaments A (region~A),
and filaments B (region~B). Black points denote the full sample,
orange points galaxies with $\log M_*/M_\odot < 10$, and dark green
points galaxies with $\log M_*/M_\odot \geq 10$. The dashed red line
at $45\degr$ indicates the isotropic expectation.}
    \label{fig:app_alignment_summary}
\end{figure*}

Figure~\ref{fig:app_alignment_summary} shows the median alignment angle
$\langle\theta\rangle$ between the projected semi-major axis and a reference
cosmic-web direction, as a function of morphological type and large-scale
environment, for the two mass bins defined in Sect.~\ref{sec:data_methods}.
Error bars are bootstrap uncertainties on the median. Filled circles indicate
subsamples with $n \geq 5$ galaxies; open triangles mark bins with $n < 5$
and should be interpreted with caution.

Concerning alignment with the cluster direction ($\theta_{\mathrm{cl}}$, top
row), the clearest signal is seen for elliptical galaxies in clusters, whose
median $\langle\theta_{\mathrm{cl}}\rangle$ lies noticeably below $45\degr$
with little dependence on stellar mass. This indicates a tendency for their
major axes to point toward the cluster centre, consistent with radial alignment
driven by tidal stretching \citep{Joachimi2015,Hirata2004}. S0 galaxies in
clusters show no comparably clear signal, while in groups the situation
partially reverses: S0s display a more pronounced offset below $45\degr$,
while ellipticals are broadly consistent with isotropy. These morphological
differences between clusters and groups are intriguing but remain tentative
given the modest sample sizes here in the PPSC. We note that the cluster/group distinction
also reflects the different nature of the two surveyed regions: region~A,
centered on the massive Perseus cluster, contains a higher proportion of
evolved clusters, whereas region~B exhibits a more diverse mixture of groups
and less evolved clusters. This environmental diversity contributes to the
observed morphological variations in alignment strength.
% \sandrine{est-ce que le distingo cluster/group est corrélé aux régions A et
% B?} --> done: région A = davantage d'amas évolués (zoom sur Perseus), région B = plus de diversité groupes/amas

Regarding alignment with the filament direction ($\theta_{\mathrm{fil}}$,
middle and bottom rows), the overall picture is one of isotropy. Most
morphological types in most environments are consistent with
$\langle\theta_{\mathrm{fil}}\rangle \approx 45\degr$ within uncertainties.
The only marginal trend appears in filaments of region~B, where spiral
galaxies show $\langle\theta_{\mathrm{fil}}\rangle$ slightly above $45\degr$,
suggesting a tendency for their major axes to lie perpendicular to the local
filament, equivalent to a spin preferentially parallel to the filament spine
in the tidal-torque framework \citep{Codis2012,Welker2014}. No such offset is
detected in region~A, which contains the massive Perseus cluster whose strong
tidal field likely dominates over any coherent filamentary signal. This
difference between the two regions should be treated with caution: subsamples
are small, projection effects are non-negligible, and a deviation of a few
degrees from $45\degr$ does not constitute a robust detection in the present
dataset.

Taken together, the radial alignment of ellipticals toward cluster centres
reinforces the picture emerging from the correlation functions: early-type
galaxies in dense environments are preferentially shaped by the local
cluster potential. The tentative perpendicular alignment of spirals in
region~B filaments hints at a spin–filament signal consistent with the
low-mass regime of the tidal-torque model \citep{Codis2012,Joachimi2015,
Castignani2022}, though its statistical significance remains limited by
the volume probed and projection effects.

\section{Discussion}
\label{sec:discussion}

We have measured intrinsic alignment signals in the PPSC using two-point
correlation functions at multiple isophotal radii for morphologically
stratified samples. The correlation functions extend to
${\sim}0.4$--$1~h^{-1}$~Mpc in comoving space, comparable to the
characteristic scales of groups and smaller than typical filament radii
(Paper~I), where tidal fields remain coherent. The use of spectroscopic
redshifts is essential for isolating physically associated galaxy pairs and
reaching this sensitivity, with direct implications for future surveys
combining deep imaging such as Euclid, LSST with spectroscopic campaigns such as
DESI \citep{DESICollaboration2016,Ivezic2019}.

The alignment signatures show minimal variation across the three isophotal
radii $R_{23}$, $R_{25}$, and $R_{27}$, with only a slight tendency for the
large-scale $\xi_{\mathrm{g}+}$ amplitude to increase with increasing
$R_{\mathrm{iso}}$, consistent with outer isophotes being more sensitive to
large-scale tidal distortions \citep{Zjupa2017}. This near-independence holds
despite 10--20\% of galaxies showing isophotal twists between consecutive radii,
indicating that local morphological perturbations do not propagate into the
statistical alignment signal at population level. Conversely, identifying these
isophotal rotations provides a useful diagnostic for systems undergoing local
tidal interactions, which could be excluded from correlation measurements in
future analyses targeting the purely large-scale tidal signal.

The morphological patterns in the correlation functions and the enrichment
analysis (Fig.~\ref{fig:env_enrichment}, Sect.~\ref{subsec:environment_analysis})
point to two physically distinct alignment mechanisms operating along
morphological lines.

For position--shape alignments, the morphological composition of SCGs mirrors
that of the full sample at nearly all environments, with only a mild excess of
early-types in clusters and groups. This near-neutrality reflects a competition
between two opposing effects: the density bias of early-types toward denser
environments enhances their contribution to $\xi_{\mathrm{g}+}$, while their
lower ellipticities suppress it, such that the two effects approximately
cancel \citep{Chisari2015}. The modest overrepresentation of early-types in
the densest environments is consistent with the well-established tendency of
pressure-supported galaxies to align radially toward the halo gravitational
potential \citep{Dressler1980,Joachimi2015}, a one-halo effect driven by
tidal interactions with the cluster potential.

For shape--shape alignments, late-type galaxies strongly dominate the SCG
population at all environments, with early-types depleted below their baseline
fraction, particularly in filaments and outskirts. The systematically higher
ellipticities of late-type SCGs ($e \approx 0.55$--$0.65$ versus $0.46$ for
the full late-type sample) are consistent with the $e^2$ weighting of $\xi_+$,
which selects the most elongated systems regardless of environment. This
behaviour is qualitatively consistent with tidal torque theory, in which
large-scale tidal fields coherently spin up disk galaxies during their
formation, producing orientations correlated across the filamentary structure
\citep{Codis2012,Welker2014}. A detection of a significant $\xi_+$ signal for
late-type galaxies is noteworthy, as intrinsic alignments in spiral populations
have historically been consistent with zero in shallower surveys
\citep{Mandelbaum2006,Joachimi2015}; the deep photometry and extensive
spectroscopy of the PPSC dataset are likely key to enabling this detection.

The early-type depletion in $\xi_+$ outside dense environments deserves
particular attention. While early-types are strongly radially aligned toward
cluster centres (Sect.~\ref{subsec:direct_alignments}), this intra-halo
alignment does not translate into coherent shape--shape correlations between
pairs at fixed separation. Their orientations reflect a superposition of
radial intra-halo alignments and large-scale tidal alignments pointing in
different directions; these can partially cancel in $\xi_+$ even when the
individual $\xi_{\mathrm{g}+}$ signal is strong. This competition is more
pronounced in underdense environments where the tidal field is less coherent
and the intra-halo signal is absent, leaving early-types with no net preferred
orientation relative to their neighbours. This provides a natural and unified
explanation for the observed morphological dichotomy.

Our results complement recent direct measurements of galaxy--filament
alignments. \citet{Muralichandran2025} studied spin--filament alignments in
a much larger sample (${\sim}32\,500$ spirals and ${\sim}19\,000$ ellipticals
from the Siena Galaxy Atlas) and found strong perpendicular alignment for
ellipticals (${\sim}13\sigma$) but only weak alignment for spirals
(${\sim}2.8\sigma$). The apparent contrast with our detection of a strong
late-type $\xi_+$ signal is not a contradiction: their estimator directly
measures the angle between a galaxy's projected axis and the nearest filament,
while $\xi_+$ measures the mutual coherence of shapes within pairs. A
population of disks with spins coherently parallel to filaments will produce
both a weak direct spin--filament signal (diluted by projection and
line-of-sight contamination) and a strong shape--shape correlation if their
ellipticities are high and their orientations are mutually consistent. This
complementarity of measurement approaches \citep{Castignani2022} underscores
the value of pair-based statistics in supercluster environments where
projection effects are substantial.

Our direct measurements of major-axis alignments with the cosmic web
(Sect.~\ref{subsec:direct_alignments}) find results broadly consistent with
isotropy, with only a marginal radial alignment of ellipticals toward cluster
centres and a tentative perpendicular alignment of spirals in region~B
filaments. These non-detections are consistent with the volume limitations
and projection effects of a single supercluster, and highlight that the
pair-based $\xi_+$ and $\xi_{\mathrm{g}+}$ estimators are more sensitive
probes of alignment in this regime.

Strongly correlated galaxies show no systematic mass enhancement relative to
the full sample. Mean stellar mass offsets range from $+0.04$ to $+0.05$~dex
and are consistent with zero within the combined uncertainties, with no
coherent trend across morphologies or correlation functions. This suggests
that stellar mass is not a primary driver of alignment strength. However, it
is important to distinguish between stellar mass and halo mass in this context.
Tidal torque theory predicts that the spin--filament alignment transition
occurs at a characteristic halo mass scale \citep{Codis2012,Welker2014}, not
a stellar mass scale; the stellar mass used as a proxy here may therefore
dilute any underlying halo-mass dependence, particularly given the scatter in
the stellar-to-halo mass relation at these masses. The comparable $\xi_+$
amplitudes across our two stellar mass bins, combined with findings from
\citet{Kraljic2020,Kraljic2021} and \citet{Barsanti2022} showing that morphology
and bulge fraction correlate more strongly with spin--filament alignments than
stellar mass alone, suggest that morphology and environment are the dominant
determinants of alignment strength in our sample, with halo mass playing a
secondary role that our stellar mass proxy is insufficient to isolate.

\section{Conclusion}\label{sec:conclusion}

We have presented the first comprehensive study of intrinsic galaxy alignments in 
the Perseus--Pisces Supercluster, combining deep low-surface-brightness CFHT 
$r$-band imaging reaching 28~mag~arcsec$^{-2}$ with the detailed three-dimensional 
cosmic web reconstruction from Paper~I. By measuring galaxy shapes at multiple 
isophotal radii ($R_{23}$, $R_{25}$, $R_{27}$) and computing their two-point 
correlation functions in comoving coordinates, we have detected significant 
intrinsic alignment signals and identified the galaxy populations that contribute 
most strongly to these correlations.

Our main findings can be summarised as follows:
\begin{itemize}
    \item Both position-shape $\xi_{\mathrm{g}+}$ and shape-shape $\xi_+$
    correlations show positive signals at separations up to $\sim$1~Mpc$/h$ 
    in comoving space, with $\xi_+$ exhibiting substantially stronger amplitudes 
    (S/N~$\sim$10 at $r \lesssim 0.1~h^{-1}$~Mpc) than $\xi_{\mathrm{g}+}$ 
    (S/N~$\sim$3 at similar scales). Strongly correlated galaxies (SCGs), defined 
    as systems with cumulative pair contributions exceeding one sigma above the mean
    in the first three comoving separation bins where $\xi$ is significantly positive, 
    represent only $\sim$12--19\% of the total population.
  
    \item Late-type galaxies strongly dominate the shape-shape correlation signal, 
    with $86\%$ of SCGs in $\xi_+(r)$ being late-types despite representing only 
    $69.3\%$ of the sample. Early-types show no significant overrepresentation in 
    position-shape correlations ($31.0 \pm 2.3\%$ of SCGs in $\xi_{\mathrm{g}+}(r)$ 
    versus their $30.7\%$ baseline fraction), though environmental enrichment becomes 
    clearer when considering their concentration near cluster and group centers. 
    This morphological segregation reflects distinct physical processes: tidal 
    stretching in dense cluster environments for early-types versus coherent tidal 
    torquing along filaments for late-types.
   
    \item Strongly correlated late-type galaxies exhibit systematically higher 
    ellipticities ($e \approx 0.55$ -- 0.65) than the full sample ($e \approx 0.46$), 
    with minimal systematic offsets in position angle. The observed ellipticity 
    enhancement ratios ($\langle e|{\rm SCG}\rangle / \langle e \rangle = 1.25$ for 
    $\xi_{\mathrm{g}+}$ and $1.73$ for $\xi_+$) are consistent with the expected 
    $e$ and $e^2$ weighting of the two estimators, confirming that alignment signals 
    arise from edge-on disk galaxies (whose angular momentum vectors are aligned with 
    the local tidal field) rather than from preferential orientations on the sky.
    
    \item Galaxies contributing strongly to correlation functions reside 
    systematically closer to both filament spines (by $\sim$ 2~Mpc for 
    late-types, $\sim$ 0.5~Mpc for early-types) and cluster/group centers (by 
    $\sim$ 1~Mpc for late-types, $\sim$ 0.5~Mpc for early-types) than the 
    full sample. This environmental segregation preserves the morphological 
    stratification documented in Paper~I, with aligned early-types concentrated 
    near clusters and groups and aligned late-types preferentially located in 
    filamentary regions where coherent accretion flows are strongest.
    
    \item Correlation functions computed at $R_{23}$, $R_{25}$, and $R_{27}$ 
    produce nearly identical profiles, with only a slight tendency for the large-scale 
    $\xi_{\mathrm{g}+}$ amplitude to increase with $R_{\mathrm{iso}}$, consistent 
    with outer isophotes being more sensitive to large-scale tidal distortions. This 
    near-independence holds despite 10--20\% of galaxies showing ellipticity changes 
    and position angle rotations between consecutive radii.
    
\end{itemize}

These results complement recent observational measurements in spiral populations 
\citep{Welker2020,Kraljic2021,Muralichandran2025} and demonstrate the value of 
combining photometric shapes with spectroscopic redshifts for upcoming surveys 
like \textit{Euclid} \citep{Laureijs2011,Mellier2025}, DESI 
\citep{DESICollaboration2016}, and LSST \citep{Ivezic2019}. The observed 
morphology-dependent patterns reinforce the need for weak lensing analyses to 
account for galaxy type and environment when modeling intrinsic alignment 
systematics, including blue/late-type galaxies.

Key limitations include modest sample size, restriction to a single supercluster 
environment, and large uncertainties for early-type subsamples in underdense regions. 
Our inability to detect direct galaxy-filament alignment angles, unlike 
\cite{Muralichandran2025}, highlights the complementary nature of statistical pair 
correlations versus direct geometric measurements \citep{Castignani2022}. Future 
work extending these measurements to diverse local superclusters with 
\textit{Euclid} first data release will assess how representative our findings 
are, though well-characterized local benchmarks remain essential for calibrating 
large statistical samples. The combination of \textit{Euclid}'s exquisite image 
quality and optimized low-surface-brightness processing \citep{EROPerseusOverview, Cuillandre2025} 
will enable shape measurements extending to $R_{29}$ and beyond, probing alignment 
signals in the outermost stellar halos where tidal coupling to the cosmic web is 
expected to be strongest. Early results from the Perseus-Pisces region observed 
during \textit{Euclid} Early Release Observations \citep{Mondelin2025} demonstrate 
the potential for such deep measurements across multiple nearby superclusters.

\bibliographystyle{aa} 
\bibliography{ref}
\begin{acknowledgements}
Based on observations obtained at the Canada-France-Hawai'i Telescope (CFHT) which is operated by the National Research Council of Canada, the Institut National des Sciences de l'Univers of the Centre National de la Recherche Scientifique of France, and the University of Hawai'i. CFHT is located on Maunakea on Hawai'i Island, a mountain of considerable cultural, natural, and ecological significance. Maunakea is a sacred site to Native Hawaiians, also known as Kānaka 'Ōiwi. Quality observations are made possible by relentless effort of the entire staff at Canada-France-Hawai'i Telescope. Based on observations obtained with MegaPrime/MegaCam, a joint project of CFHT and CEA/DAPNIA. \\
This research is based in part on data collected at Subaru Telescope,
which is operated by the National Astronomical Observa-
tory of Japan. Pan-STARRS is a project of the Institute for Astronomy of the University of Hawaii, and is supported
by the NASA SSO Near Earth Observation Program under
grants 80NSSC18K0971, NNX14AM74G, NNX12AR65G,
NNX13AQ47G, NNX08AR22G, 80NSSC21K1572, and by
the State of Hawaii.
This research used the facilities of the Canadian Astron-
omy Data Centre operated by the National Research Council
of Canada with the support of the Canadian Space Agency.
Additionally, we are grateful for the computing resources of
the Digital Research Alliance of Canada and the CANFAR
(Canadian Advanced Network for Astronomical Research)
Science Portal for enabling the analysis in this paper.
\\
This research has made use of the NASA/IPAC Extragalactic Database (NED),
which is operated by the Jet Propulsion Laboratory, California Institute of Technology, under contract with the National Aeronautics and Space Administration. We also acknowledge the use of the HyperLEDA database, the tools and catalogs provided by the CDS, and the publicly available data from the 2MRS, SDSS, and FASHI surveys, which were essential for the analyses presented in this work.
\\
We thank Florence Durret for her insightful comments and suggestions that improved this paper.
\\
This work was also supported by the CEA Research Funding (Contrat formation par la recherche) program. 
\end{acknowledgements}

\begin{appendix}

\section{Comparison \texttt{TreeCorr} and manual computation}\label{app:comparisonTreecorr}

\begin{figure}
    \centering
    \includegraphics[width=0.49\textwidth]{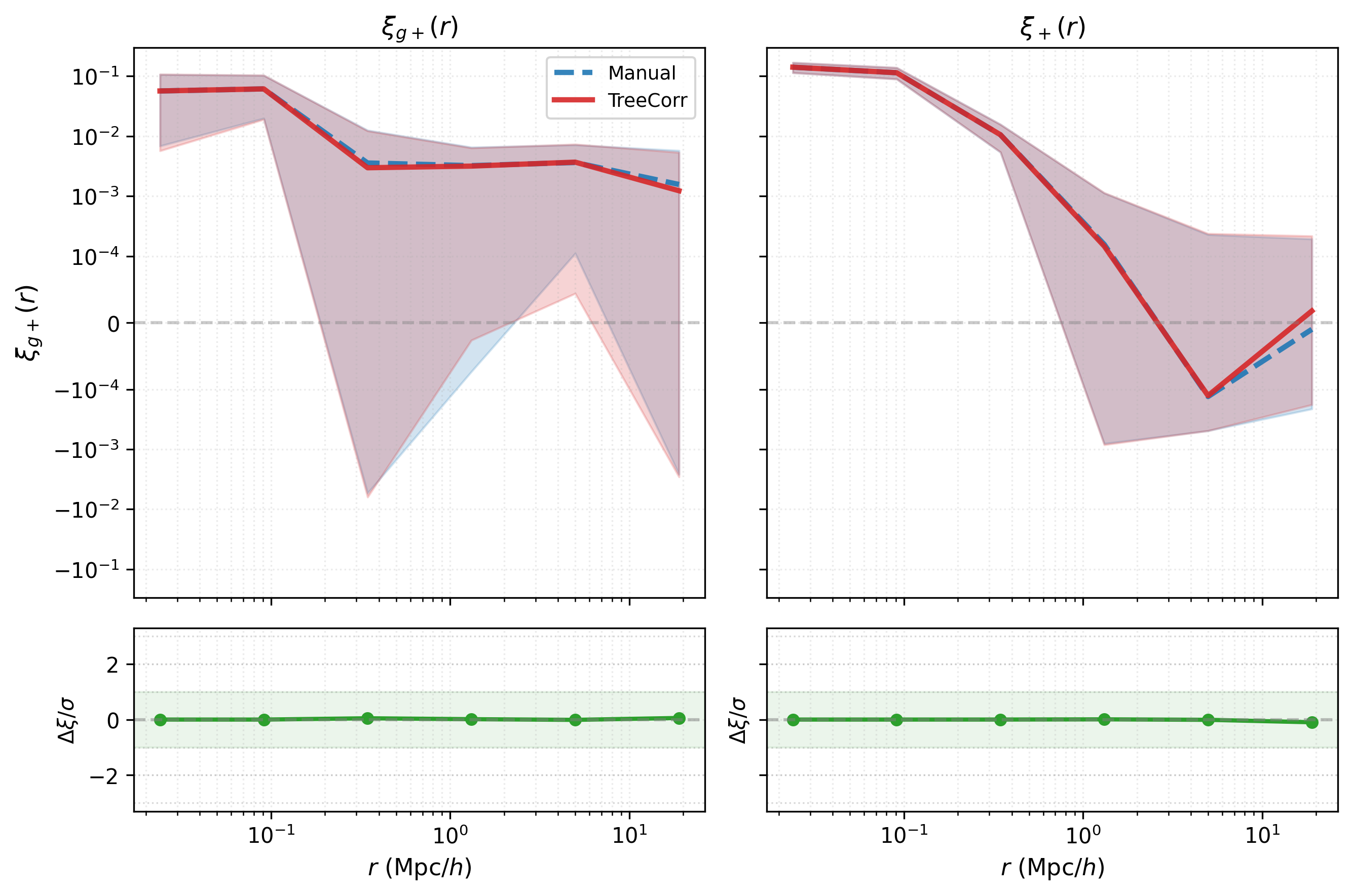}
    \caption{Validation of manual shear correlation functions (blue dashed) against 
    \texttt{TreeCorr} (red solid) at $R_{25}$. Top: $\xi_{g+}$ (left) and $\xi_+$ (right) 
     in comoving coordinates. Shaded regions show  errors. Bottom: Normalized differences ($\Delta\xi/\sigma$). All measurements agree 
    within $\pm1\sigma$ (green band), confirming our manual implementation.}
    \label{fig:manualtreecorrR25}
\end{figure}

Figure~\ref{fig:manualtreecorrR25} validates our manual correlation implementations against \texttt{TreeCorr} \citep{Jarvis2015} at $R_{25}$. Both methods employ identical jackknife resampling for error estimation. The lower panels display normalized differences ($\Delta\xi/\sigma_{\rm combined}$), showing that all measurements agree within $\pm1\sigma$ (green band).

\section{Morphology of galaxies}\label{app:morpho}

\begin{figure*}
    \centering
    \includegraphics[width=\textwidth]{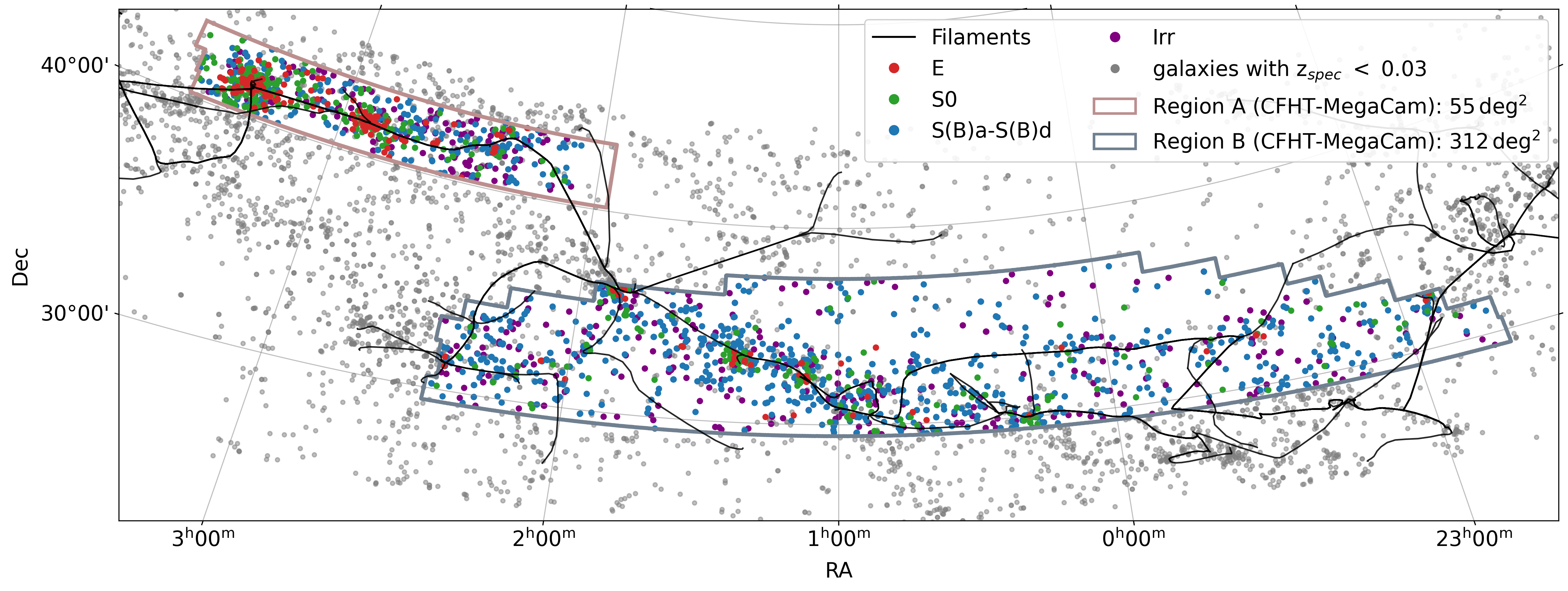}
    \caption{Projected distribution of galaxies in the PPSC colour-coded by
morphological type. Grey points show the full galaxy catalogue, coloured points
indicate galaxies with morphological classifications, and black curves trace the
projected 3D cosmic-web skeleton from Paper~I.}
    \label{fig:footprint_with_morpho}
\end{figure*}

Figure~\ref{fig:footprint_with_morpho} shows the projected distribution of galaxies
in the PPSC colour-coded by morphological type. This figure provides a visual
complement to the quantitative environmental trends discussed in
Sect.~\ref{subsec:environment_analysis}. Early-type galaxies are preferentially
located in the densest regions, in particular around cluster and group centres,
whereas late-type galaxies are more broadly distributed along filamentary
structures and in the outskirts. This projected morphology--environment
segregation is consistent with the environmental classification inherited from
Paper~I and supports the interpretation that the alignment signal depends not
only on morphology, but also on where each morphological population resides
within the cosmic web.

The figure also highlights the different spatial sampling of the two observed
regions. Region~A contains a more prominent
concentration of early-type galaxies associated with the dense cluster
environment. Region~B shows a more extended distribution of late-type galaxies
along lower-density filamentary structures. These differences motivate the
separate discussion of direct filament alignments in regions A and B in
Sect.~\ref{subsec:direct_alignments}.

\section{Morphology-dependent correlation functions}
\label{app:morpho_correlations}

\begin{figure*}
    \centering
    \includegraphics[width=\textwidth]{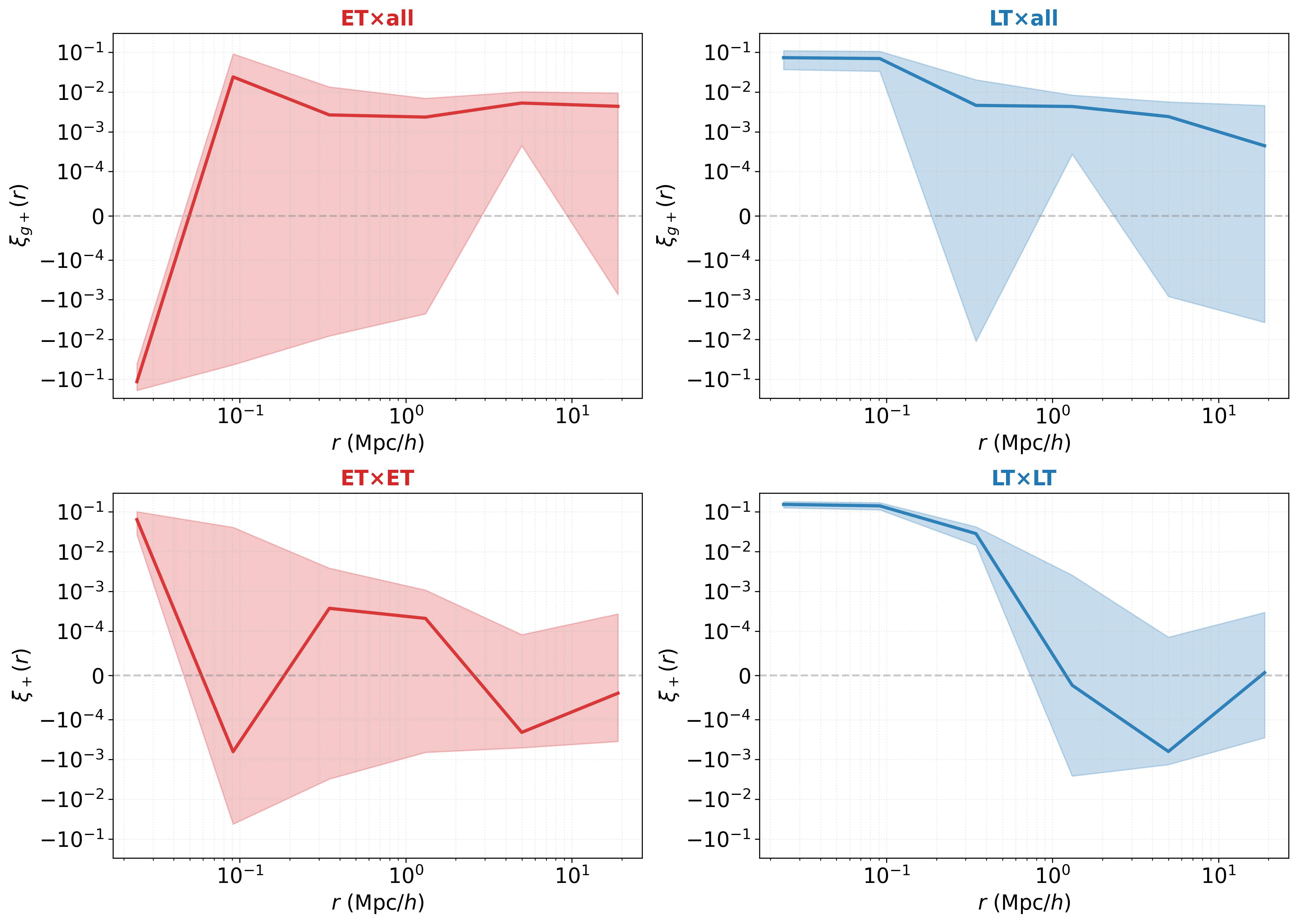}
    \caption{Morphology-dependent correlation functions at $R_{25}$. Upper row: 
             position-shape correlation $\xi_{\mathrm{g}+}(r)$ for ET×all (left, red) 
             and LT×all (right, blue). Lower row: shape-shape correlation $\xi_+(r)$ 
             for ET×ET (left, red) and LT×LT (right, blue). Shaded regions indicate 
             bootstrap uncertainties. The dashed grey line marks zero correlation.}
    \label{fig:cross_morphology}
\end{figure*}

Figure~\ref{fig:cross_morphology} presents correlation functions computed separately 
for different morphological cross-correlations  measured at $R_{25}$. The upper row shows $\xi_{\mathrm{g}+}(r)$ 
for ET×all (left) and LT×all (right), measuring position-shape correlations between 
galaxies of a specific morphological type (late or early type) as shape tracers and the full galaxy population as density tracers. The lower row shows 
$\xi_+(r)$ for ET×ET (left) and LT×LT (right), measuring shape-shape correlations 
within each morphological type.

These decomposed correlation functions reveal that late-type galaxies clearly 
dominate the intrinsic alignment signal, while early-type correlations suffer from 
large uncertainties that prevent robust conclusions. For ET×all, the $\xi_{\mathrm{g}+}(r)$ 
measurement shows large uncertainties across all scales, making it difficult to 
assess the contribution of early-types to the position-shape signal. Similarly, 
the ET×ET shape-shape correlation $\xi_+(r)$ exhibits uncertainties comparable to 
or exceeding the signal amplitude, precluding a meaningful detection.

By contrast, LT×all shows a clear positive $\xi_{\mathrm{g}+}(r)$ signal extending at least 
to $r \sim 0.1~h^{-1}$~Mpc, with amplitudes reaching ${\sim}9\times10^{-2}$ at small 
separations before declining toward the baseline. This confirms that late-type 
galaxies contribute significantly to the position-shape correlation measured for 
the full sample (Fig.~\ref{fig:corrtot}).

The LT×LT shape-shape correlation $\xi_+(r)$ provides the most striking result: 
a robust positive signal with amplitudes of ${\sim}0.1$--$0.2$ at 
$r \lesssim 0.1~h^{-1}$~Mpc, remaining significantly positive out to 
${\sim}0.4~h^{-1}$~Mpc. This demonstrates that the strong $\xi_+$ signal detected 
in the full sample (Fig.~\ref{fig:corrtot}) is carried almost entirely by the 
late-type population. The extended correlation scale (${\sim}0.4~h^{-1}$~Mpc) 
indicates that late-type shape alignments persist well beyond the one-halo regime, 
consistent with coherent tidal torquing along filamentary structures where late-types 
preferentially reside (Sect.~\ref{subsec:environment_analysis}).

These findings confirm the morphological dichotomy identified in 
Sect.~\ref{subsec:dominant_galaxies}: shape-shape alignment signals are dominated 
by late-type galaxies, whose high ellipticities (reflecting preferentially edge-on 
configurations) and coherent orientations along the cosmic web produce strong 
mutual shape correlations. The inability to robustly detect ET×ET correlations 
does not imply that early-type galaxies are unaligned, but rather reflects the 
smaller sample size, lower mean ellipticities, and the potential cancellation 
between radial alignments toward cluster centers and large-scale tidal alignments 
discussed in Sect.~\ref{sec:discussion}.
\end{appendix}

\end{document}